\documentclass[aps,prc,twocolumn,amsmath,10pt,superscriptaddress,floatfix,nofootinbib]{revtex4-2}

\usepackage{epsfig,amssymb,amsfonts,amsmath,mathtools,bm,color,xcolor,graphicx}

\usepackage{hyperref}
\hypersetup{pdftex,colorlinks=true,linkcolor=blue,citecolor=blue,menucolor=black,urlcolor=blue,filecolor=blue}

\graphicspath{{Figures/}}

\newcommand{\itp}{\affiliation{CAS Key Laboratory of Theoretical Physics, Institute of Theoretical Physics,\\
Chinese Academy of Sciences, Beijing 100190, China}}

\newcommand{\ucas}{\affiliation{School of Physical Sciences, University of Chinese Academy of Sciences, Beijing 100049, China}}

\newcommand{\peng}{\affiliation{Peng Huanwu Collaborative Center for Research and Education, Beihang University, Beijing 100191, China}}

\newcommand{\darm}{\affiliation{Institut für Kernphysik, Technische Universität Darmstadt, 64289 Darmstadt, Germany}}

\newcommand{\emmi}{\affiliation{ExtreMe Matter Institute EMMI and Helmholtz Forschungsakademie Hessen für FAIR (HFHF),\\ GSI Helmholtzzentrum für Schwerionenforschung GmbH, 64291 Darmstadt, Germany}}

\begin{document}

\title{Neutron Scattering off One-Neutron Halo Nuclei in Halo Effective Field Theory}

\author{Xu Zhang}\email{zhangxu@itp.ac.cn}
\itp

\author{Hai-Long Fu}\email{fuhailong@itp.ac.cn}
\itp\ucas

\author{Feng-Kun Guo}\email{fkguo@itp.ac.cn}
\itp\ucas\peng

\author{Hans-Werner Hammer}\email{Hans-Werner.Hammer@physik.tu-darmstadt.de}
\darm\emmi

\begin{abstract}

Neutron scattering off neutron halos can provide important information about the internal structure of nuclei close to the neutron drip line.
In this work, we use halo effective field theory to study the $s$-wave scattering of a neutron and the spin-parity $J^P=\frac{1}{2}^+$ one-neutron halo nuclei 
$^{11}\rm Be$,  $^{15}\rm C$, and $^{19}\rm C$ at leading order. In the $J=1$ channel, the only inputs to the Faddeev equations are their
one-neutron separation energies. 
In the $J=0$ channel, the neutron-neutron scattering length and
the two-neutron separation energies of $\rm ^{12}Be$, $\rm ^{16}C$ and $\rm ^{20}C$ enter as well. The numerical results show
that the total $s$-wave cross sections in the $J=1$ channel at threshold are of the order of a few barns. 
In the $J=0$ channel, these cross sections are of the order of a few barns for $n$-$^{11}\rm Be$ and $n$-$^{19}\rm C$ scattering,
and about 60 $\rm mb$ for the $n$-$^{15}\rm C$ scattering. 
The appearance of a pole in $p\cot\delta$ close to zero  in all three cases
indicates the existence of a virtual Efimov state close to threshold in each of the $^{12}\rm Be$,  $^{16}\rm C$, and $^{20}\rm C$ systems. Observation of this pole would confirm the presence of Efimov physics in halo nuclei.
The dependence of the results on the neutron-core scattering length is also studied.

\end{abstract}

\maketitle

\section{Introduction}

Over the last four decades, various nuclei have been discovered where the valence neutron(s) have a large probability to distribute in the classically forbidden region outside the range of the core potential~\cite{Zhukov:1993aw,Jensen:2004zz,jonson:2004,Tanihata:2013jwa}. 
To a good approximation, these nuclei can be described as a compact,
structureless core surrounded by a halo of valence neutron(s). 
The unusual size of the halo nuclei can be viewed as a consequence of quantum mechanical tunnelling of the halo neutrons out of the core potential.
Understanding the structure of halo nuclei provides a window to fundamental aspects of the nuclei along the neutron drip line.

Effective field theory (EFT) provides a powerful tool to explore nuclear systems where separation of scales exists~\cite{Weinberg:1990rz,Weinberg:1991um}.
Depending on the desired resolution scale, different EFTs for nuclear phenomena have been constructed (see Refs.~\cite{Beane:2000fx,Bedaque:2002mn,Epelbaum:2005pn,Epelbaum:2008ga,Bogner:2009bt,Machleidt:2011zz,Hammer:2012id,Holt:2014hma,Papenbrock:2015mdi,Hammer:2017tjm,Hammer:2019poc} for reviews of these efforts). 
In an EFT, one can construct the most general Lagrangian involving low-energy degrees of freedom while the short-distance physics can be described by a derivative expansion of local interactions.
Physical observables can be expanded in powers of the short-distance over large-distance scales. EFT allows a systematic and controlled approach to 
investigating nuclear interactions in low-energy processes. The separation of scales in halo nuclei implies that one can use an
EFT for halo nuclei (the so-called Halo EFT), which uses the core and the halo nucleons as degrees of freedom~\cite{Bertulani:2002sz,Bedaque:2003wa,Higa:2008dn}.  Their interaction is described by contact terms expanded in powers of $Q/\Lambda$, where $Q$ is a general low-energy momentum scale, while $\Lambda$ is the breakdown scale of Halo EFT. It is set by the lowest scale of physics not explicitly included, such as the pion mass $m_{\pi}\sim140$~MeV or the momentum corresponding to a core excitation. 
In Halo EFT, the core and valence neutron(s) are treated as point-like particles. The finite size of the core enters only in higher-order corrections.
 
Halo EFT has been used to study a variety of processes. The theory was first applied to study the shallow $p$-wave neutron-alpha 
resonance~\cite{Bertulani:2002sz,Bedaque:2003wa} and the $s$-wave alpha-alpha resonance~\cite{Higa:2008dn}. Then, it was extended to investigate three-body
systems. For example, the structure of two-neutron halo nuclei and the relavance of the Efimov effect~\cite{Efimov:1970zz,Efimov:1970xx,Amado:1971dlx,Amado:1972yss,Federov:1994cf,NFJG,Amorim:1997mq,Braaten:2004rn,Naidon:2016dpf}
were studied at leading order (LO) in Refs.~\cite{Canham:2008jd,Hagen:2013xga,Acharya:2013aea,Ji:2014wta,Pang:2023prr}
and next-to-leading order (NLO) in Refs.~\cite{Canham:2009xg,Vanasse:2016hgn}. 
Halo EFT has also been used to study the electromagnetic reactions of neutron and proton halo nuclei including range corrections and higher-order electromagnetic interactions (see, {\it e.g.}., Refs~\cite{Hammer:2011ye,Rupak:2011nk,Rupak:2012cr,Acharya:2013nia,Zhang:2013kja,Zhang:2015ajn,Zhang:2019odg,Gobel:2023wtb}).
Reviews of the applications of Halo EFT can be found in Refs.~\cite{Hammer:2017tjm,Hammer:2022lhx}.

The halos $^{11}\rm Be$,  $^{15}\rm C$ and $^{19}\rm C$ with spin-parity $J^P=\frac{1}{2}^+$ all have a core with $J^P=0^+$.
The lowest excitation energies $E_c^*$ of the cores $^{10}\rm Be$,  $^{14}\rm C$ and $^{18}\rm C$ are 3.368, 6.094 and 1.62~MeV according to the Triangle Universities Nuclear Laboratory (TUNL) database~\cite{TUNL,Ajzenberg-Selove:1987,Ajzenberg-Selove:1991rsl,Tilley:2004zz}, respectively.
Moreover, the one-neutron separation energies $B_{\sigma}$ of $^{11}\rm Be$,  $^{15}\rm C$ and $^{19}\rm C$ are 0.502, 1.218 and 0.58~MeV from the atomic mass evaluation AME2020~\cite{Huang:2021nwk,Wang:2021xhn}, respectively.
A rough estimate of the expansion parameters $Q/\Lambda\sim\sqrt{B_{1n}/{E_c^*}}$  in Halo EFT gives about $0.39$ for $^{11}\rm Be$, $0.45$ for $^{15}\rm C$ and $0.60$ for $^{19}\rm C$, respectively. These estimates of the expansion parameter $Q/\Lambda$ are consistent with
an analysis of electromagnetic reactions, as discussed in Refs.~\cite{Hammer:2011ye,Rupak:2012cr,Acharya:2013nia,Fernando:2015jyd,Moschini:2019tyx}. 

Neutron scattering off halo nuclei can provide important information about the internal structure of the nuclei in the neutron drip line region.
Such experiments may be performed with radioactive beams of the halo nuclei~\cite{Zhan:2010rzb, Gales:2010dsu, Motobayashi:2010edb, Thoennessen:2010hdv, Zhou:2022pxl} on a deuterium target.
The reactions can be computed from the neutron-halo and proton-halo scattering amplitudes. 
In principle, they may also be performed directly at facilities with neutron beams,
such as the China Spallation Neutron Source (CSNS)~\cite{Wei:2009aa,Chen:2016aa}, if appropriate beams of the halo nuclei become available.  

In this work, we study the interaction of neutron and the spin-parity $J^P=\frac{1}{2}^+$ one-neutron halo nuclei 
$^{11}\rm Be$,  $^{15}\rm C$ and $^{19}\rm C$ using Halo EFT.
There are two channels in the $s$-wave scattering, corresponding to the total spin $J=1$ and $J=0$. 
Our calculation is performed at LO. At this order, the only inputs to the Faddeev equations are the one-neutron separation energies of 
one-neutron halo nuclei in total spin $J=1$ channel, as well as the two-neutron scattering length and the two-neutron separation energies of $\rm ^{12}Be$, $\rm ^{16}C$ and $\rm ^{20}C$ in the total spin $J=0$ channel. Corrections from higher orders can be estimated as the theoretical uncertainty.

The paper is organized as follows. In Section~\ref{sec:Lagrangiam}, we introduce the effective Lagrangian for the interactions
necessary for calculating the scattering of neutron and $^{11}\rm Be$,  $^{15}\rm C$ and $^{19}\rm C$.  
Section~\ref{sec:TwoBody} describes how the relevant two-body interactions emerge from the effective Lagrangian.
The three-body interactions are studied in Section~\ref{sec:ThreeBody}.
Numerical results and discussions are presented in Section~\ref{sec:Results}. Finally, we summarize our work in Section~\ref{sec:Sum}. Some technical details are given in the appendices.

\section{Effective Lagrangian}\label{sec:Lagrangiam}

In Halo EFT, the relevant degrees of freedom are the core nucleus ($c$) and the valence nucleon(s). At LO, the two-body subsystems $nn$ and $nc$ are described
by zero-range interactions. 
The effective Lagrangian can be written as a sum of one-body, two-body and three-body contributions,
\begin{equation}
\mathcal{L}=\mathcal{L}_1+\mathcal{L}_2+\mathcal{L}_3\,.
\end{equation}
The one-body Lagrangian is
\begin{equation}
\mathcal{L}_1 
= \vec{n}^{\dagger} \left( i\partial_0 + \frac{\nabla^2}{2m_n} \right) \vec{n} 
+ c^{\dagger} \left( i\partial_0 + \frac{\nabla^2}{2m_c} \right) c\,, 
\end{equation}
where $c$ is a scalar field for the core nuclei $^{10}\rm Be$,  $^{14}\rm C$ and $^{18}\rm C$ with a mass $m_c$, the masses are taken from Refs.~\cite{Huang:2021nwk,Wang:2021xhn} as given in Table~\ref{Tab:InputMass}, 
and $\vec{n}$ represents a two-component spinor field of the valence neutron $\vec{n}=(n_\uparrow~~n_\downarrow)$ with a
mass $m_n$. The mass of neutron is taken to be $m_n=939.57$~MeV~\cite{ParticleDataGroup:2022pth}.
\begin{table}[tb]
\caption{The mass of the core nuclei. The data are taken from AME2020~\cite{Huang:2021nwk,Wang:2021xhn}. 
}
\begin{ruledtabular}
\setlength{\tabcolsep}{5mm}{
\begin{tabular}{lcccc}
Nuclei&  $\rm ^{10}Be$ & $\rm ^{14}C$ & $\rm ^{18}C$  \tabularnewline
\hline 
$m_c$ (MeV) &  $9327.5$ & $13043.9$ & $16791.8$  
\end{tabular}}
\end{ruledtabular}
\label{Tab:InputMass}
\end{table}
If the effective ranges are positive,
the two-body Lagrangian involving the $s$-wave $nn$ and $nc$ interactions to NLO can be written as~\cite{Bedaque:1997qi,Bedaque:1998mb,Bedaque:1999ve,Hammer:2011ye,Acharya:2013nia}
\begin{align}
\label{Eq:LTwo}
\mathcal{L}_2=&\, s^{\dagger}\left[\Delta_s -\left( i\partial_0 + \frac{\nabla^2}{4m_n} \right) \right]s  \nonumber \\
&+ \sigma_i^{\dagger}\left[\Delta_{\sigma}  -\left( i\partial_0 + \frac{\nabla^2}{2m_{\sigma}} \right) \right]\sigma_i \nonumber \\
&-g_sC_{1/2\alpha,1/2\beta}^{00}\left[s^{\dagger}n_{\alpha}n_{\beta}+\rm{H.c.}\right]-g_{\sigma}\!\left[\sigma_i^{\dagger}n_ic+\rm{H.c.}\right], 
\end{align}
where $s$ is a scalar dimer field for the two-neutron system, $\vec{\sigma}=(\sigma_\uparrow~~\sigma_\downarrow)$ is a two-component spinor field for $^{11}\rm Be$,  $^{15}\rm C$ and $^{19}\rm C$
with a mass $m_{\sigma}=m_c+m_n$, and $C_{1/2\alpha,1/2\beta}^{00}$ is a Clebsch-Gordan coefficient.
The parameters $\Delta_s$, $\Delta_{\sigma}$, $g_s$ and $g_{\sigma}$ can be determined by experimental data. At LO,
$\Delta_{s,\sigma}$ and $g_{s,\sigma}$ are not independent and only
the combinations $g_{s,\sigma}^2/\Delta_{s,\sigma}$ enter into
observables.

The three-body interaction Lagrangian $\mathcal{L}_3$ can be constructed in terms of the $s$-wave two-neutron dimer and the core~\cite{Hagen:2013xga,Vanasse:2016hgn},
which is needed for the renormalization in the total spin $J=0$ channel at LO. It can be written as
\begin{equation}
\mathcal{L}_3 
= g_{s}^2 {D_0}(sc)^{\dagger}(sc)\,, \label{eq:C0}
\end{equation}
with ${D_0}$ a three-body parameter.

\section{Two-body interactions}\label{sec:TwoBody}

\begin{figure}[t]
\begin{center}
\includegraphics [scale=0.4] {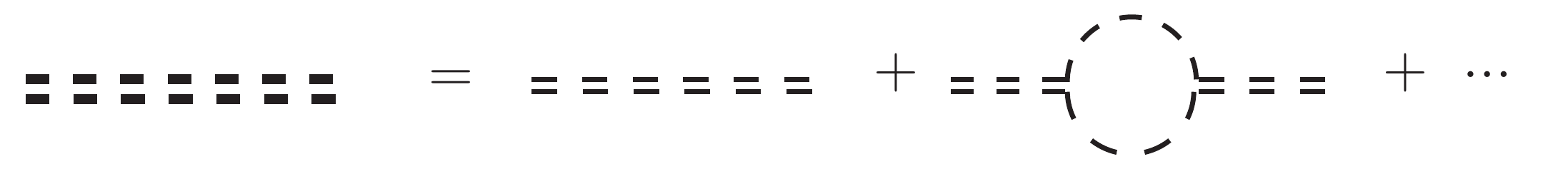}\\
\includegraphics [scale=0.4] {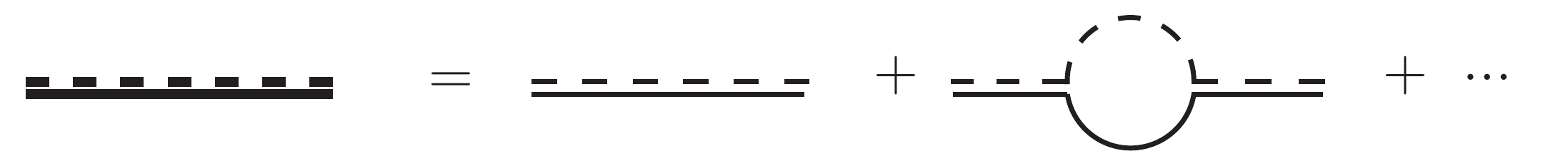}
\caption{Dimer propagators at LO from the Lagrangian in Eqs.~\eqref{Eq:LTwo}.
The double-dashed line and dashed-solid line are the $nn$ and $nc$ propagators, respectively.
The bare dimer propagator $i/{\Delta_{s,\sigma}}$ is dressed by an infinite number of bubble diagrams.
The dashed and solid lines represent the neutron and the core nucleus, respectively.}
\label{fig:DimerDress}
\end{center}
\end{figure}
At LO, the bare dimer propagators $i/{\Delta_{s,\sigma}}$ are dressed by an infinite number of bubble diagrams as shown in Fig.~\ref{fig:DimerDress}.
Using the Feynman rules, the full $nn$ and $nc$ dimer propagators consisting of the geometric series represented in Fig.~\ref{fig:DimerDress} can
be written as
\begin{align}
&iD_s(p_0,{p})=  \nonumber \\  
&\frac{-i}{-\Delta_s+m_ng_s^2/(2\pi)\left[-\Lambda+\sqrt{m_n[{p}^2/(4m_n)-p_0-i\epsilon]}\right]}\,,\nonumber \\
&iD_{\sigma}(p_0,{p})=\nonumber \\ 
&\frac{-i}{-\Delta_{\sigma}+\mu_{nc}g_{\sigma}^2/(2\pi)\left[-\Lambda+\sqrt{2\mu_{nc}[{p}^2/(2m_{\sigma})-p_0-i\epsilon]}\right]}\,,
\end{align}
where $p_0$ and $p$ are the energy and the magnitude of the three-momentum of the dimer field, respectively. 
$\mu_{nc}=m_nm_c/(m_n+m_c)$ is the reduced mass of the $nc$ two-body system, and 
$\Lambda$ is an arbitrary momentum scale in the power divergence subtraction (PDS)  scheme~\cite{Kaplan:1998tg,Kaplan:1998we}.

For the $nn$ two-body system, the parameters $\Delta_s$ and $g_s$ can be matched to the experimentally known scattering length $a_s$,
\begin{equation}
a_s^{-1}=\frac{2\pi}{m_ng_s^2}\Delta_s+\Lambda\,,
\end{equation}
and we take the value $a_s=-18.6$~fm~\cite{Chen:2008zzj}.
The full renormalized $nn$ dimer propagator can then be written as
\begin{align}
iD_s(p_0,{p})=\frac{2\pi}{m_ng_s^2}\frac{-i}{-1/a_s+\sqrt{m_n[{p}^2/(4m_n)-p_0-i\epsilon]}}\,.
\end{align}
For a negative scattering length $a_s$, the pole of the propagator is on the second Riemann sheet of the complex energy plane 
and the residue at the pole is negative,
\begin{equation}
Z_s^{-1}=\frac{m_n^2g_s^2a_s}{4\pi}\,.
\end{equation}

For the $nc$ two-body system, the parameters $\Delta_{\sigma}$ and $g_{\sigma}$ are connected to the scattering length $a_{\sigma}$ by
\begin{equation}
a_{\sigma}^{-1}=\frac{2\pi}{\mu_{nc}g_{\sigma}^2}\Delta_{\sigma}+\Lambda\,,
\end{equation}
and the full renormalized $nc$ dimer propagator can then be written as
\begin{align}
iD_{\sigma}(p_0,{p})=\frac{2\pi}{\mu_{nc}g_{\sigma}^2}\frac{-i}{-1/a_{\sigma}+\sqrt{2\mu_{nc}[{p}^2/(2m_{\sigma})-p_0-i\epsilon]}}\,.
\end{align}
The $nc$ dimer propagator has a pole at
\begin{equation}
p_0-\frac{{p}^2}{2m_{\sigma}}=-\frac{1}{2\mu_{nc}a_{\sigma}^2}
=-B_\sigma,
\end{equation}
where $B_{\sigma}$ is the separation energy between the core and the valence neutron of the one-neutron halo nucleus.
The input parameters for the $nc$ two-body interaction are given in Table~\ref{Tab:InputA}.
\begin{table}[tb]
\caption{EFT inputs for the $nc$ two-body interactions. The data are taken from AME2020~\cite{Huang:2021nwk,Wang:2021xhn}. 
}
\begin{ruledtabular}
\setlength{\tabcolsep}{5mm}{
\begin{tabular}{lcccc}
Halo nuclei&  $\rm ^{11}Be$ & $\rm ^{15}C$ & $\rm ^{19}C$  \tabularnewline
\hline 
$B_{\sigma}$ (MeV) &  $0.502$ & $1.218$ & $0.58$  \tabularnewline
$a_{\sigma}$ (fm) &  $6.741$ & $4.27$ & $6.142$ 
\end{tabular}}
\end{ruledtabular}
\label{Tab:InputA}
\end{table}
For positive scattering length $a_{\sigma}$, the pole of the propagator is on the first Riemann sheet of the complex square root 
and the residue at the pole is positive,
\begin{equation}
Z_{\sigma}^{-1}=\frac{\mu_{nc}^2g_{\sigma}^2a_{\sigma}}{2\pi}\,.
\end{equation}
In our calculation, we stay at LO in the power counting of Halo EFT, and the effective range corrections arising from the 
kinetic terms of the dimers in Eq.~(\ref{Eq:LTwo}) are not included~\cite{Bethe:1949yr,Bedaque:1999vb,Bedaque:2002yg,Griesshammer:2004pe}.

\section{Three-body interaction}\label{sec:ThreeBody}

In this work, we are interested in the $s$-wave scattering of neutron and the one-neutron halo nuclei $^{11}\rm Be$, $^{15}\rm C$ and $^{19}\rm C$ in Halo EFT.
The Efimov effect appears in three-body systems when the scattering length of the two-body subsystem is much larger than the range of the forces~\cite{Efimov:1970zz,Efimov:1970xx,Amado:1971dlx,Amado:1972yss,NFJG,Federov:1994cf,Braaten:2004rn}.
The condition for the appearance of the Efimov state in halo nuclei was first studied in Ref.~\cite{Federov:1994cf}.
The Efimov effect in $^{12}\rm Be$ and $^{20}\rm C$ was studied in the renormalized zero-range model in Ref.~\cite{Amorim:1997mq}. 
In Refs.~\cite{Canham:2008jd,Canham:2009xg} it was investigated in Halo EFT at LO and NLO. Taking the values of the $nc$ two-body binding energy $B_{\sigma}$ from the TUNL database~\cite{Ajzenberg-Selove:1987,Tilley:2004zz,Ajzenberg-Selove:1991rsl}, 
the numerical calculations at LO indicate that only $^{20}\rm C$ with the $n$-$^{18} \rm C$ bound state energy $B_{\sigma}=(0.162\pm 0.112)$~MeV has a possible excited Efimov state, with an energy less than
14~keV in Ref.~\cite{Amorim:1997mq} and 7~keV in Ref.~\cite{Canham:2008jd} below the $n$-$^{19}\rm C$ threshold.
The shift in the excited state binding energy is found less than 0.5~keV at NLO in Ref.~\cite{Canham:2009xg}.
However, recent evaluation in AME2020~\cite{Huang:2021nwk,Wang:2021xhn} 
finds $B_{\sigma}=(0.58\pm 0.09)$~MeV for the $n$-$^{18}\rm C$ interaction, which
does not support an excited Efimov state in $^{20}\rm C$.
Experiments on neutron scattering off halo nuclei provide important information about the internal structure of the nuclei,
and can put constraints on the possible existence of excited Efimov states in these nuclei.

The radius of halo nuclei is much larger than the radius of their core, and the core can be assumed as point-like.
This separation of scales allows for a description of the low-energy properties of halo nuclei using contact interactions between the halo neutrons and between the neutrons and the core.
The $T$-matrix for scattering of a neutron and a one-neutron halo nucleus will be calculated at LO in Halo EFT, 
by solving the corresponding Faddeev integral equation. The integral equation was first derived in Ref.~\cite{Skorniakov} for the 
neutron-deuteron system and is identical to the result given in Ref.~\cite{Faddeev:1961} if contact interactions are considered. Including only two-body interactions,
the numerical results for $s$-wave observables in
the three-body system of two neutrons and the core nucleus in the total spin $J=0$ channel, where three-body bound states appear,
suffer from a peculiar cutoff dependence. This behavior was already observed in  Refs.~\cite{Danilov:1961,Minlos:1962,Danilov:1963} for the 
$J=1/2$ neutron deuteron system and also occurs in the three-boson system.
This cutoff dependence can be removed by introducing a  
three-body force that runs log-periodically with the cutoff~\cite{Bedaque:1998kg,Bedaque:1999ve,Mohr:2005pv}. 
Fixing the three-body force from a single three-body observable,
all other observables can be predicted at LO.
In our calculations, the three-body forces in the $J=0$ channel are fixed with the $^{12}\rm Be$,  $^{16}\rm C$ and $^{20}\rm C$
ground state two-neutron separation energies, and the log-periodic behavior of the three-body forces is also observed as expected (see below).

\begin{widetext}
\subsection{Total spin $J=1$ channel}

The $T$-matrix for $l$-th partial wave $n\sigma$ scattering   in the total spin $J=1$ channel is given by the integral equation displayed in Fig.~\ref{fig:SpinOne}.
Since the spin of the core nucleus of interest is zero, the total spin of the two neutrons should be one for this channel. Such a configuration is not forbidden by the Pauli exclusion principle because one of the neutrons is bound inside the neutron halo nucleus. Thus the identical fermions are not localized in a point. 
Three-body contact terms without derivatives, however, are forbidden by the exclusion principle. Indeed one finds that no three-body force contribution at LO is needed to achieve a cutoff independent solution. 
The integral equation for this channel can be written as 
 \begin{figure*}[t]
\begin{center}
\includegraphics [scale=0.4] {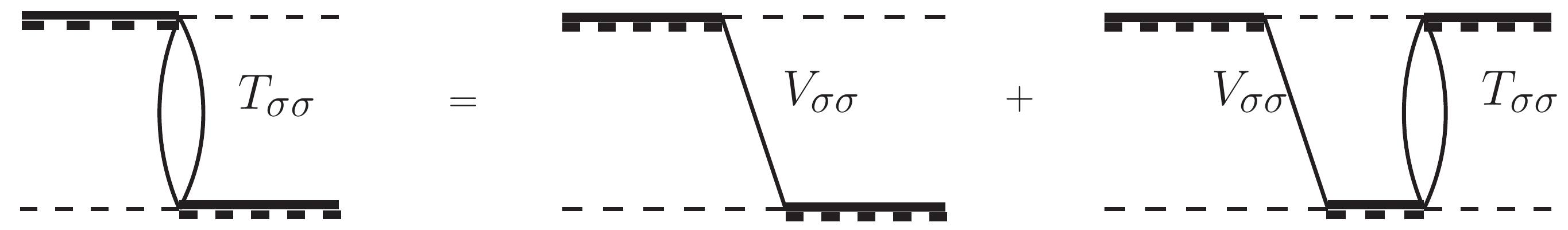}
\caption{Faddeev equation for the $n\sigma$ scattering in the total spin $J=1$ channel at LO. The notations are the same as those in Fig.~\ref{fig:DimerDress}.}
\label{fig:SpinOne}
\end{center}
\end{figure*}
\begin{equation}
\label{Eq:FaddSpinOne}
iT_{\sigma\sigma}^{\,l}(k,p,E)=iV_{\sigma\sigma}^{\,l}(k,p,E)-\int_0^{\Lambda}\frac{q^2dq}{2\pi^2}V_{\sigma\sigma}^{\,l}(k,q,E)\,Z_{\sigma}^{-1}D_{\sigma}\!\left(E-\frac{q^2}{2m_n},q\right) iT_{\sigma\sigma}^{\,l}(q,p,E)\,,
\end{equation}
where $p$ and $k$ are the magnitudes of the incoming and outgoing momenta of the spectator neutron in the 3-body center-of-mass frame, respectively, and the upper index $l$ denotes the partial wave. 
The total nonrelativistic on-shell energy is 
\begin{equation}
\label{Eq:OnShellE}
E=\frac{p^2}{2 \mu_{n\sigma}}-\frac{1}{2\mu_{nc}a_{\sigma}^2}
=\frac{p^2}{2 \mu_{n\sigma}}
-B_\sigma\,,
\end{equation}
where 
\begin{equation}
\label{Eq:nsigmared}
\mu_{n\sigma}=\frac{m_nm_{\sigma}}{m_n+m_{\sigma}}\,,
\end{equation}
is the $n\sigma$ reduced mass.
Notice that when all external legs are on-shell, the momenta $p$ and $k$ are equal.

The $l$-th partial wave projection of the core-exchange potential can be written as
\begin{align}
\label{Eq:CEP}
V_{\sigma\sigma}^{\,l}(k,q,E)&=Z_{\sigma} g_{\sigma}^2\cdot\frac{1}{2}\int_{-1}^{+1}\frac{P_l(\cos\theta)\, d\cos\theta } {E-k^2/(2m_n)-q^2/(2m_n)-(\vec{q}+\vec{k})^2/(2m_c)+i\epsilon}\,, \nonumber \\
&=Z_{\sigma} g_{\sigma}^2\cdot\frac{m_c}{qk}\,Q_l\!\left(\frac{m_cE-m_\sigma/(2m_n)(q^2+k^2)+i\epsilon}{qk}\right)\,,
\end{align}
where $\theta$ is the angle between $\vec q$ and $\vec k$, the Legendre polynomial of the second kind with complex argument is
\begin{align}
Q_l(z)=\frac{1}{2}\int_{-1}^{+1}\frac{P_l(\cos\theta)\, d\cos\theta } {z-\cos\theta}\,,
\end{align}
and the analytic expressions of the first few $Q_l(z)$ are given in Appendix~\ref{sec:legendre}.

\subsection{Total spin $J=0$ channel}

\begin{figure*}[t]
\begin{center}
\includegraphics [scale=0.4] {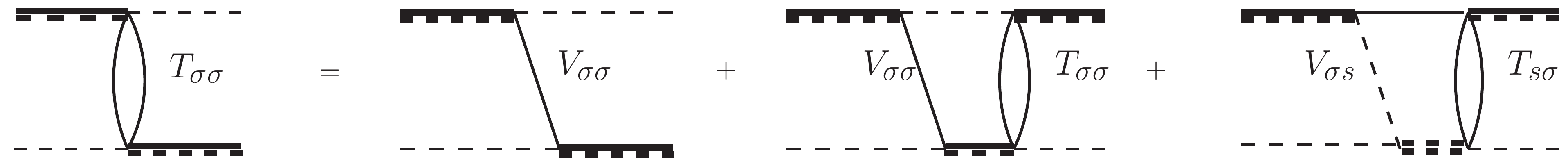}
\includegraphics [scale=0.4] {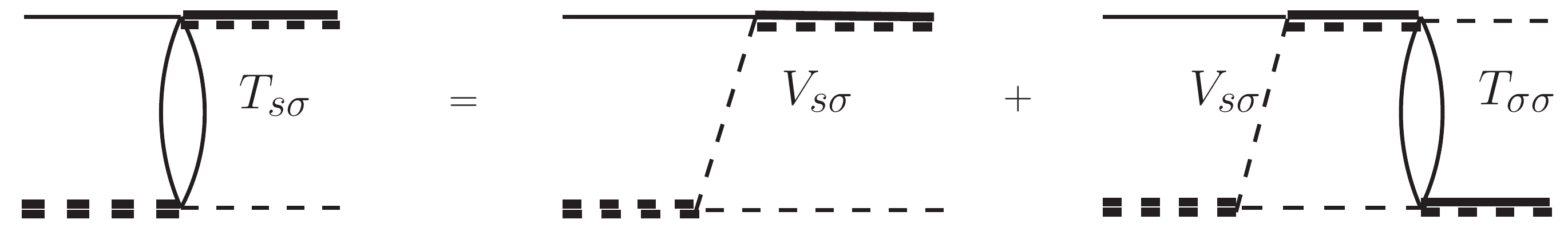}~~~~~~~~~~~~~~~~~~~~~~~~~~~~~~~~~~~~~~~~~~
\caption{Faddeev equation for the $n\sigma$ scattering in the total spin $J=0$ channel at LO. The notations are the same as those in Fig.~\ref{fig:DimerDress}.}
\label{fig:SpinZero}
\end{center}
\end{figure*}

The $T$-matrix for the $l$-th partial wave $n\sigma$ scattering   in the total spin $J=0$ channel is given by the integral equation shown in Fig.~\ref{fig:SpinZero}.
The integral equation for this channel can be written as 
\begin{align}
\label{Eq:FaddSpinZero}
&\begin{pmatrix}
iT_{\sigma\sigma}^{\,l}(k,p,E)\\
iT_{s\sigma}^{\,l}(k,p,E)
\end{pmatrix}
=
\begin{pmatrix}
-iV_{\sigma\sigma}^{\,l}(k,p,E)\\
i2V_{s\sigma}^{\,l}(k,p,E)
\end{pmatrix}  
\nonumber \\
&-\int_0^{\Lambda}\frac{q^2dq}{2\pi^2}
\begin{pmatrix}
-V_{\sigma\sigma}^{\,l}(k,q,E)&~~2V_{\sigma s}^{\,l}(k,q,E)\\
2V_{s\sigma}^{\,l}(k,q,E)& ~~0
\end{pmatrix}
\begin{pmatrix}
Z_{\sigma}^{-1}D_{\sigma}(E-\frac{q^2}{2m_n},q)&0\\
0& Z_s^{-1}D_{s}(E-\frac{q^2}{2m_c},q)
\end{pmatrix}
\begin{pmatrix}
iT_{\sigma\sigma}^{\,l}(q,p,E)\\
iT_{s\sigma}^{\,l}(q,p,E)
\end{pmatrix}\,.
\end{align}
The $l$-th partial wave projection of the core-exchange potential has been given in  Eq.~\eqref{Eq:CEP}.
The $l$-th partial wave projection of the neutron-exchange potential can be written as
\begin{align}
\label{Eq:NEP}
V_{\sigma s}^{\,l}(k,q,E)&=(\sqrt{Z_{\sigma}Z_s} g_{\sigma}g_s)\cdot\frac{1}{2}\int_{-1}^{+1}\frac{P_l(\cos\theta) d\cos\theta}{E-k^2/2m_n-q^2/2m_c-(\vec{q}+\vec{k})^2/2m_n+i\epsilon} \,,\nonumber \\
&=(\sqrt{Z_{\sigma}Z_s} g_{\sigma}g_s)\cdot\frac{m_n}{qk}Q_l\left(\frac{m_nE-k^2-m_\sigma/(2m_c)q^2+i\epsilon}{qk}\right)\,,
\end{align}
\end{widetext}
and 
\begin{align}
V_{s\sigma }^{\,l}(k,q,E)&=V_{\sigma s}^{\,l}(q,k,E).
\end{align}
For this channel, a three-body force is required for renormalization at LO.  For clarity, however, the three-body force terms are omitted in Fig.~\ref{fig:SpinZero} and Eq.~(\ref{Eq:FaddSpinZero}).
Their contribution will be included in the calculations below.  

The integral equations in the $J=0$ and $J=1$ channels are solved numerically. Details of the numerical solution method are given in Appendix~\ref{sec:numerical}.

\section{Numerical Results and Discussion}\label{sec:Results}

At LO of Halo EFT, the only input parameter in the $J=1$ channel is the neutron-core $s$-wave scattering length or the one-neutron separation energy of the one-neutron-halo nucleus. The values for the $\rm ^{11}Be$, $\rm ^{15}C$ and $\rm ^{19}C$ nuclei are listed in Table~\ref{Tab:InputA}. In addition, to understand the dependence of the results from the variation of the $nc$ scattering length $a_{\sigma}$, its values will be varied by $\pm0.5$~fm.
In the $J=0$ channel, two additional parameters are needed to fix the interactions: the two-neutron scattering length $a_s$ and the two-neutron separation energies $B_{2n}$ of $\rm ^{12}Be$, $\rm ^{16}C$ and $\rm ^{20}C$ to fix the three-body force. We use the values given in the AME2020~\cite{Huang:2021nwk,Wang:2021xhn}
as shown in Table~\ref{Tab:InputB}.
\begin{table}[t!]
\caption{Two-neutron separation energies $B_{2n}$. 
The data are taken from AME2020~\cite{Huang:2021nwk,Wang:2021xhn}.}
\begin{center}
\setlength{\tabcolsep}{3.5mm}{
\begin{tabular}{cccc}
\hline 
Halo nuclei & $\rm ^{12}Be$ & $\rm ^{16}C$ & $\rm ^{20}C$ \tabularnewline
\hline 
\hline 
$B_{2n}$ (MeV) &   $3.672$ & $5.468$& $3.560$ \tabularnewline
\hline 
\end{tabular}}
\end{center}
\label{Tab:InputB}
\end{table}

\subsection{Cutoff dependence in the $J=0$ channel}\label{sec:CounterTerm}

\begin{figure}[t]
\begin{center}
\includegraphics [scale=0.55] {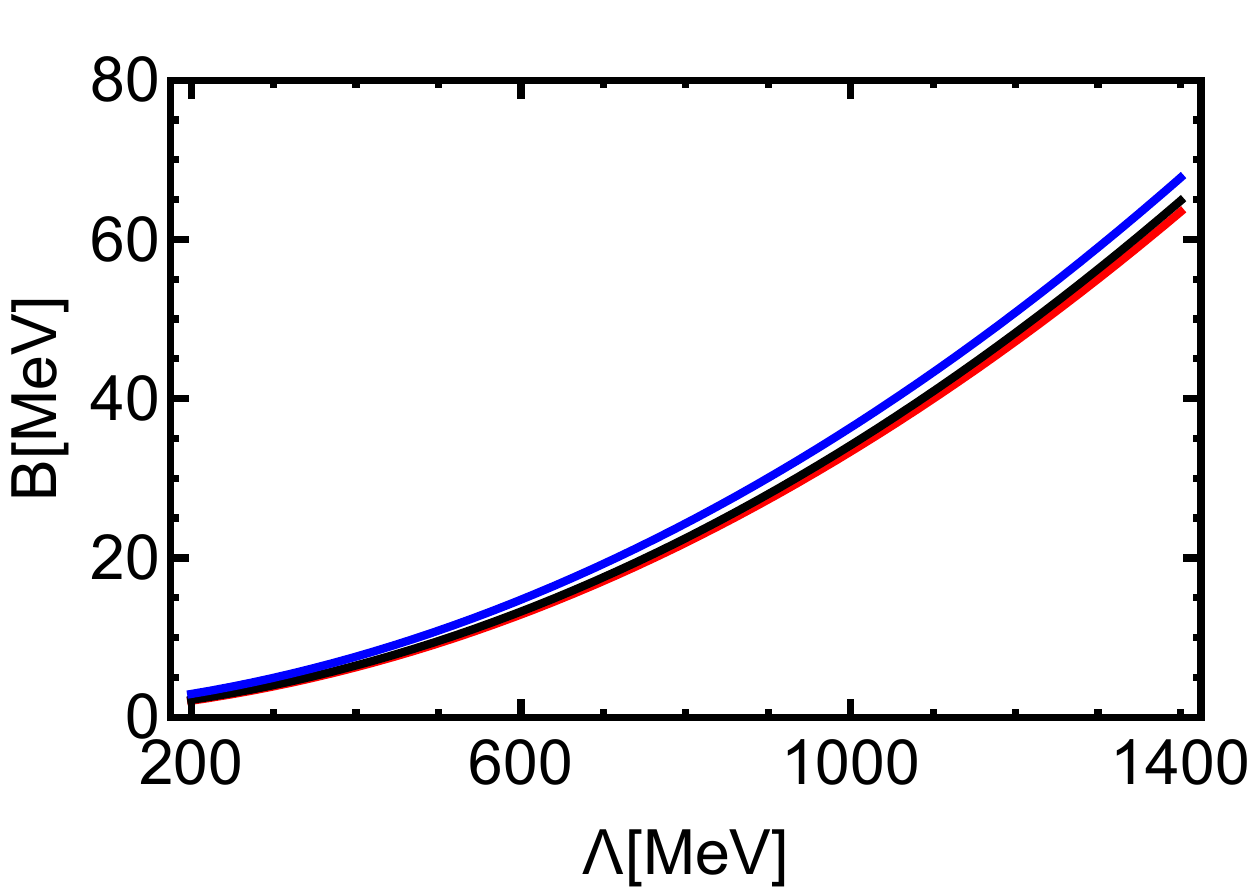}
\caption{Cutoff dependence of the three-body binding energy in the $J=0$ channel without a three-body counterterm. 
The red, blue and black lines correspond to the bound states generated from the $nn^{10}$Be, $nn^{14}$C and $nn^{18}$C three-body interactions, respectively.}
\label{fig:BindEn}
\end{center}
\end{figure}
Using a sharp cutoff $\Lambda$ to regularize the Faddeev equation in Eq.~\eqref{Eq:FaddSpinZero} 
and omitting the three-body force, 
the cutoff dependence of the three-body binding energies in the $J=0$ channel are shown in Fig.~\ref{fig:BindEn}.
One can see that the energies have a strong cutoff dependence. Asymptotically, they grow with $\Lambda^2$.
In addition, the binding energies are insensitive to the masses of different nuclei
for the given scattering lengths $a_{\sigma}$. 
This fact can be expected as the core nuclei considered here are much heavier than the neutron, and thus the core kinetic energies are much smaller than that of the neutron.\footnote{When the core nucleus is much heavier than the nucleon, one may construct an EFT treating the core nucleus as static at LO, similar to the heavy quark effective theory~\cite{Isgur:1989vq} and heavy baryon chiral perturbation theory~\cite{Jenkins:1990jv,Bernard:1995dp}.
}

\begin{figure*}[tbhp]
\begin{center}
\includegraphics [scale=0.46] {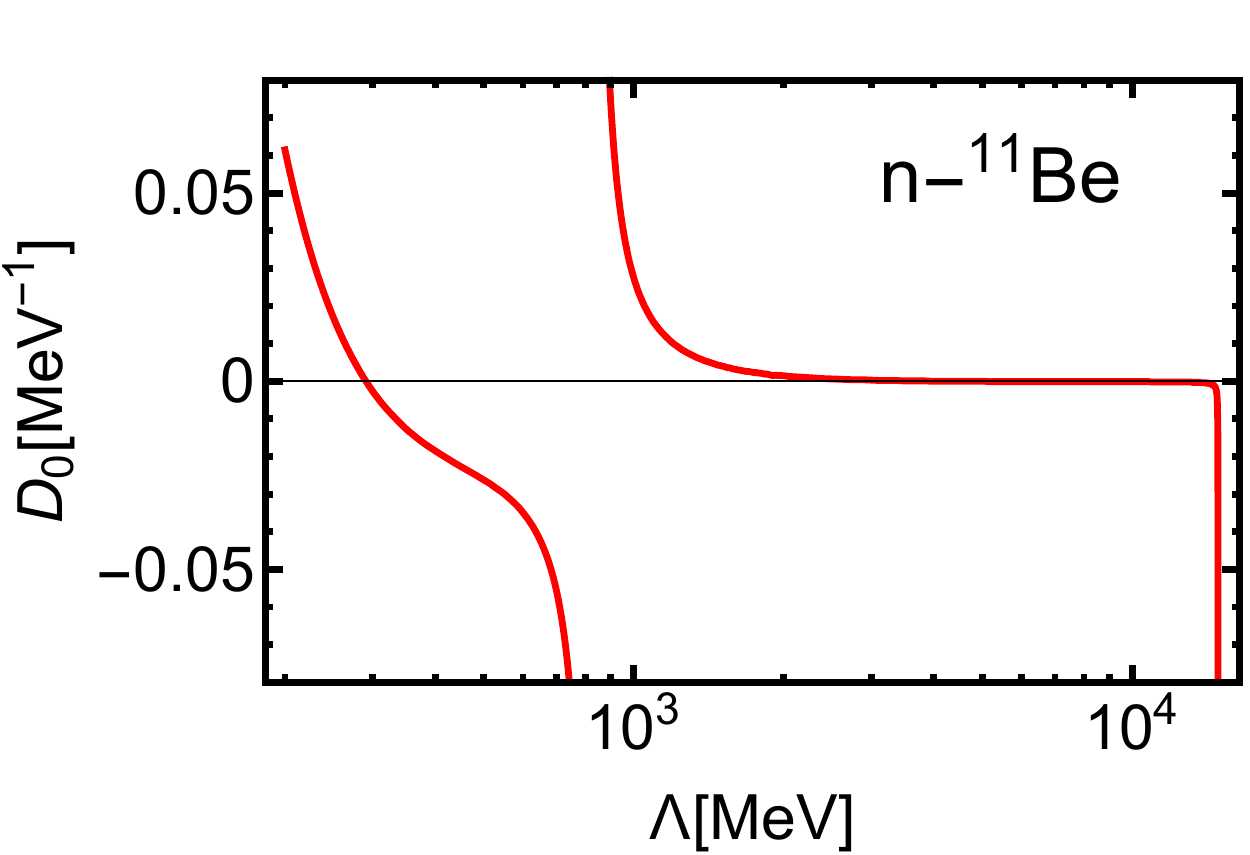}
\includegraphics [scale=0.46] {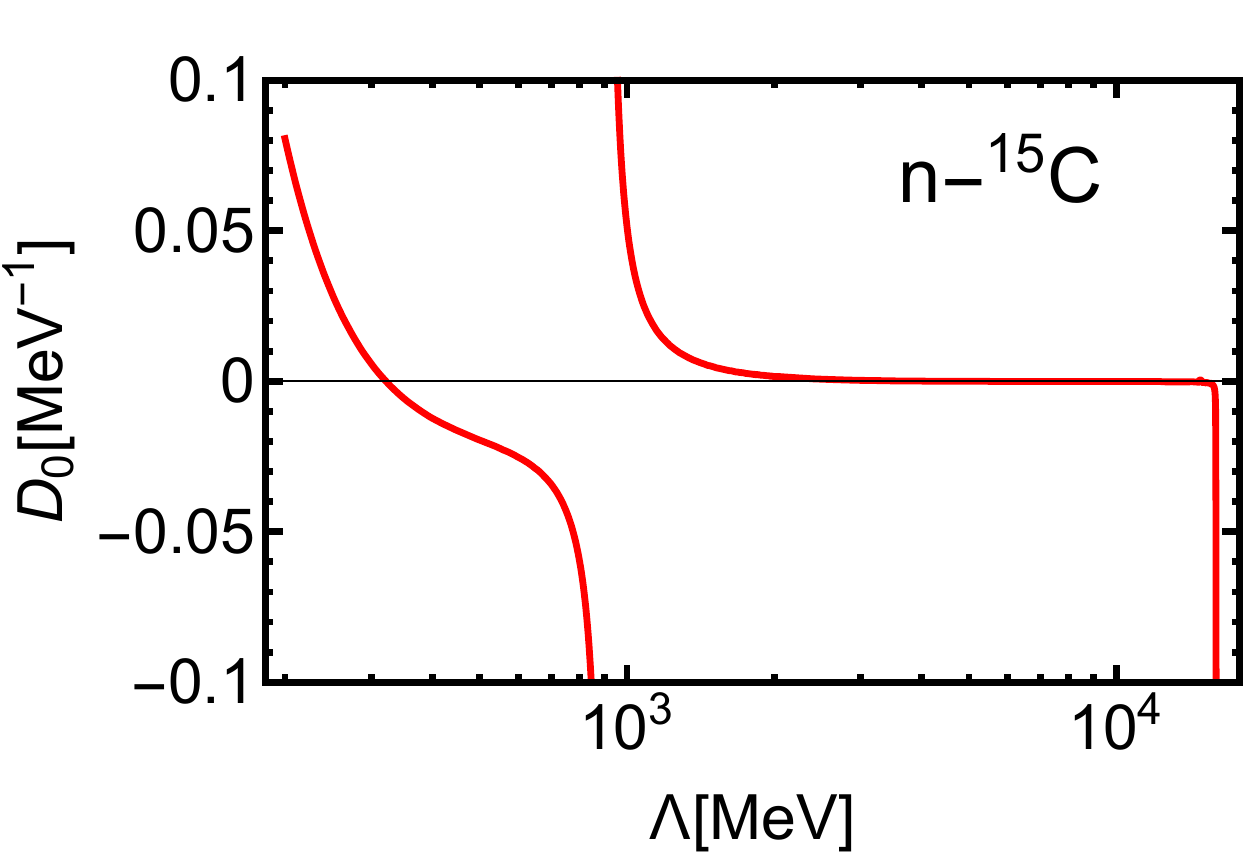}
\includegraphics [scale=0.46] {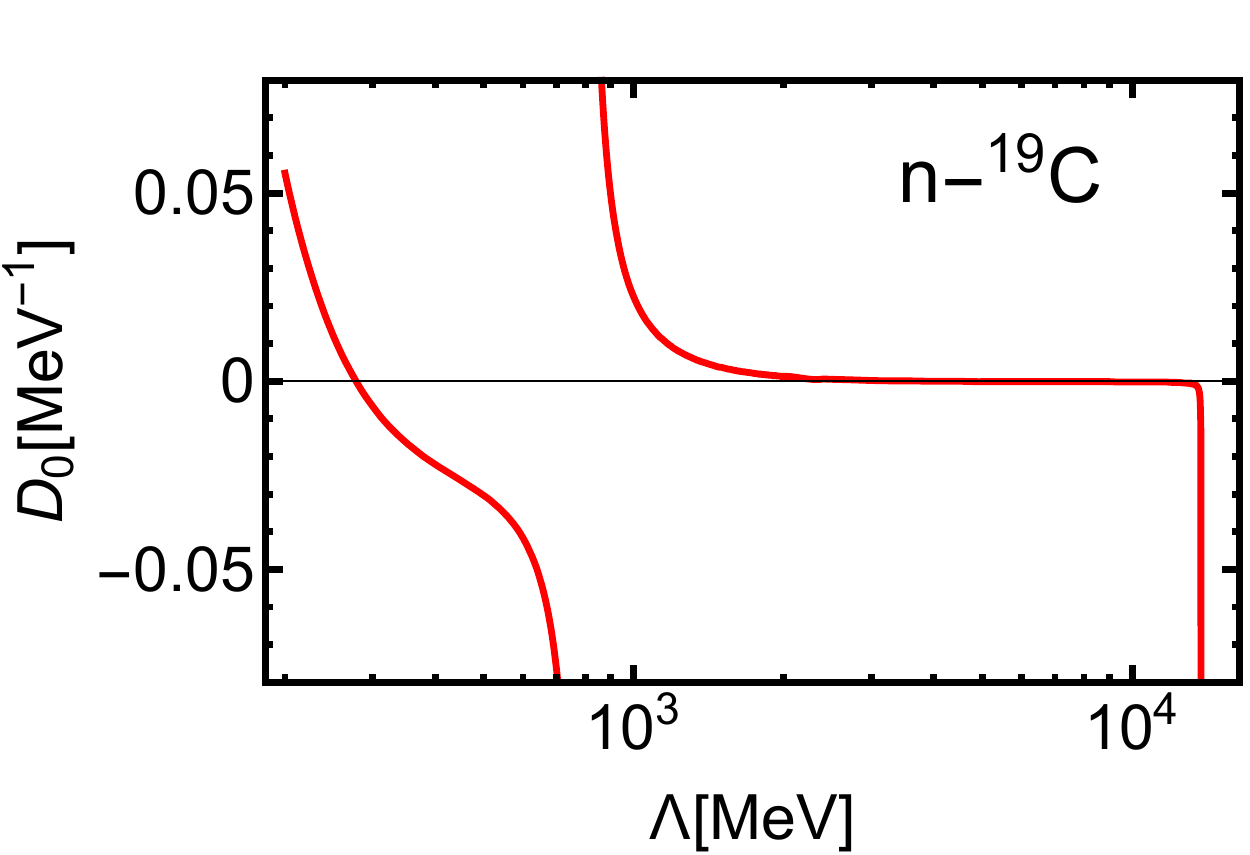}
\caption{The three-body parameter $D_0$ as a function of the cutoff $\Lambda$. 
$D_0$ is tuned to reproduce the ground state two-neutron separation energy for each of $^{12}\rm Be$ (left),  $^{16}\rm C$ (middle) and $^{20}\rm C$ (right).
}
\label{fig:ThreeBodyForce}
\end{center}
\end{figure*}

Following Refs.~\cite{Bedaque:1998kg,Canham:2008jd},
one may add a three-body counterterm to the Faddeev equation~\eqref{Eq:FaddSpinZero} to cancel this cutoff dependence,
\begin{equation}
V_{ss}~(k,p,E)=Z_sg_s^2D_0.
\end{equation}
The counterterm parameter $D_0$ is tuned to reproduce the ground state two-neutron separation energy for each of $^{12}\rm Be$,  $^{16}\rm C$ and $^{20}\rm C$. Once this renormalization is 
done, all other three-body observables, such as scattering
cross sections are also cutoff-independent.
The cutoff dependence of the three-body parameter $D_0$ is shown in Fig.~\ref{fig:ThreeBodyForce}.
One can see that the three-body parameter $D_0$ has a quasi-periodic behavior in $\ln\Lambda$.  Similar to the three-boson case~\cite{Bedaque:1998kg,Bedaque:1998km}, the function $\Lambda^2 D_0$ is log-periodic in $\Lambda$~(see Ref.~\cite{Hammer:2017tjm} for a detailed investigation of this behavior for different core masses).

\begin{figure*}[bthp]
\begin{center}
\includegraphics [width=0.32\textwidth] {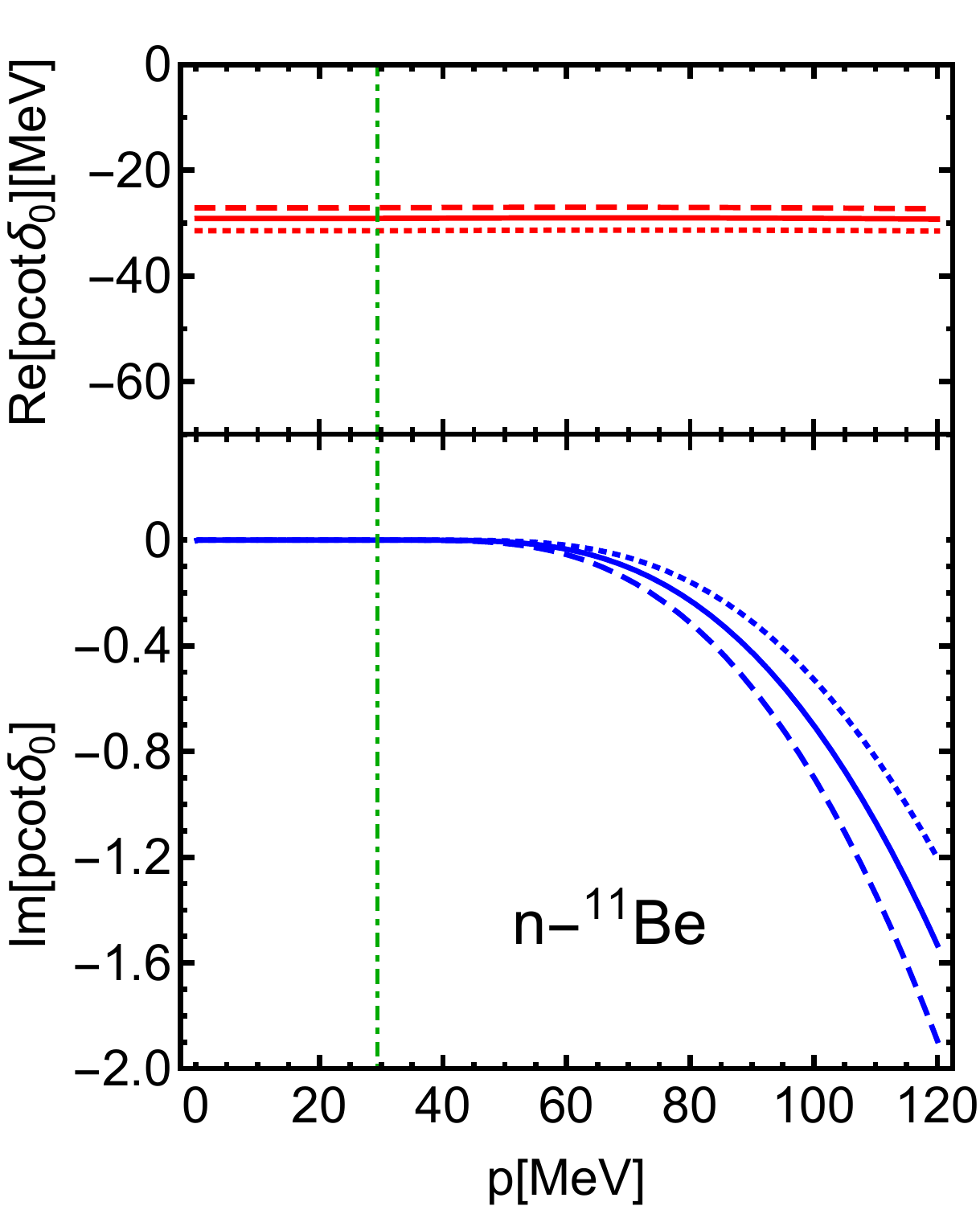}
\includegraphics [width=0.32\textwidth] {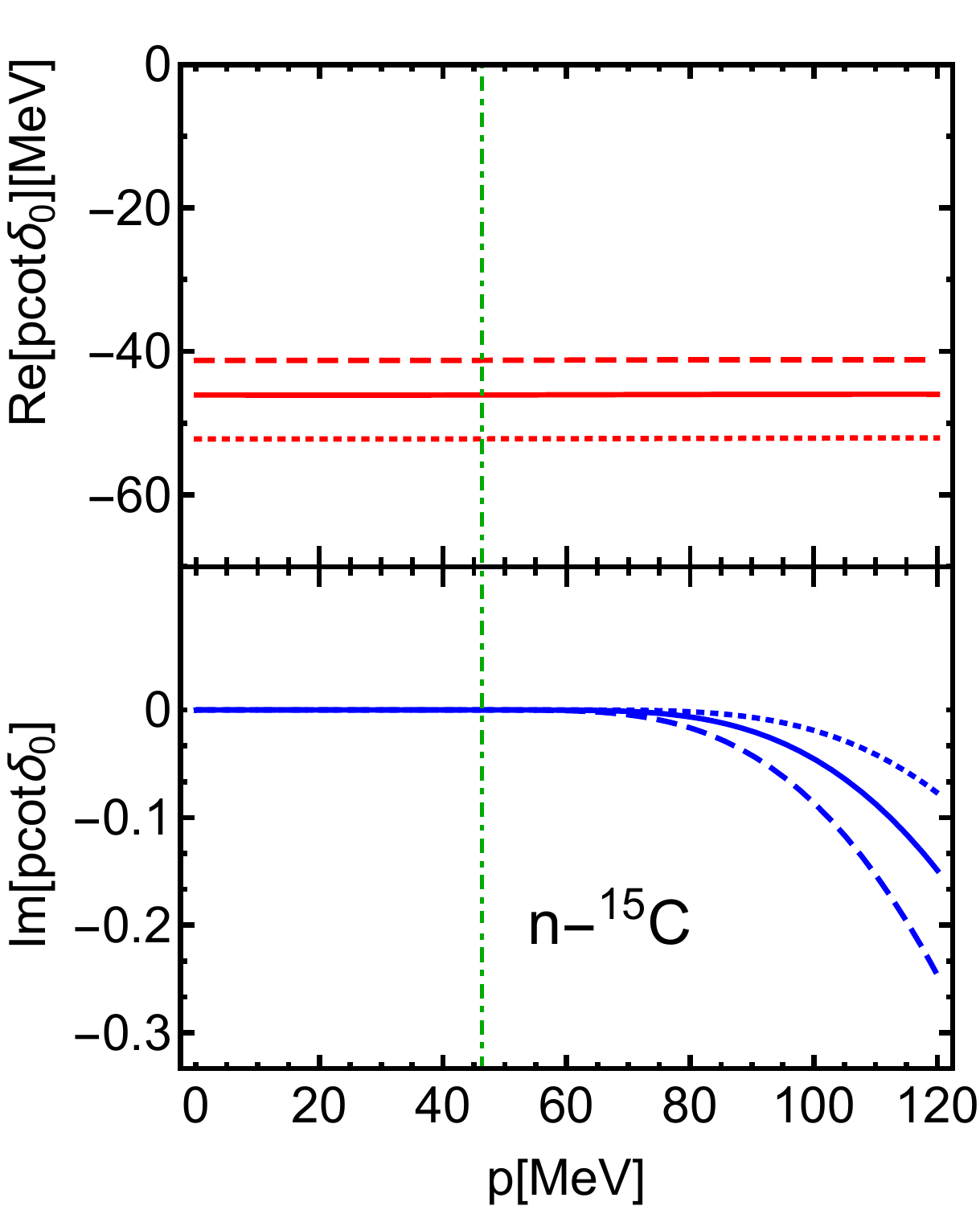}
\includegraphics [width=0.32\textwidth] {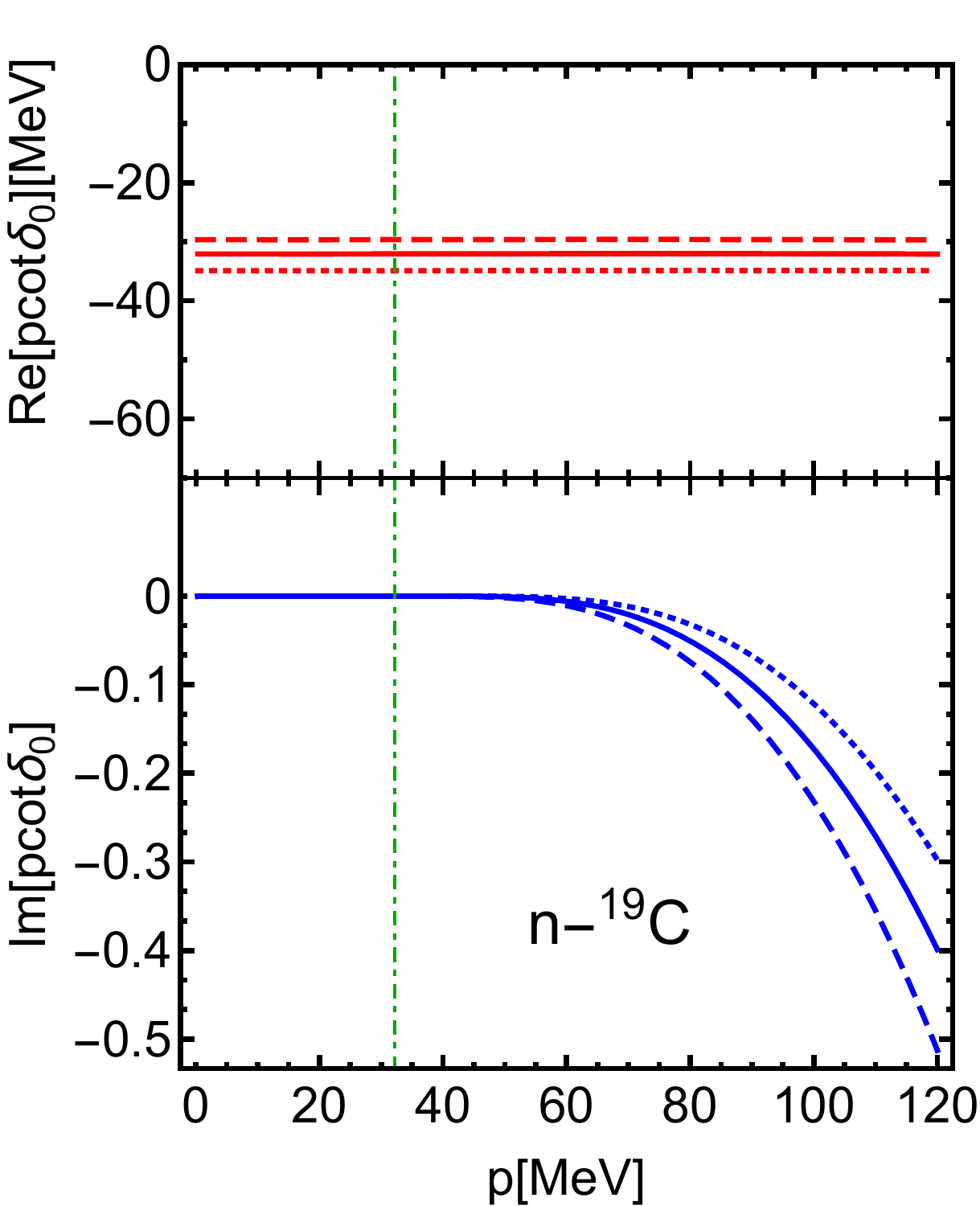}\\
\caption{Real and imaginary parts of $p\cot\delta_0(p)$ for the $s$-wave scattering of neutron and one-neutron halo nuclei in the $J=1$ channel. The left, middle and right panels correspond to the $n\text{-}^{11}\rm Be$, $n\text{-}^{15}\rm C$ and $n\text{-}^{19}\rm C$ scattering processes, respectively. 
The solid lines represent the results using the scattering lengths $a_\sigma$ in Table~\ref{Tab:InputA} as inputs, while the dashed/dotted curves correspond to $a_\sigma \pm 0.5$~fm, respectively. 
The vertical dot-dashed lines in the plots indicate 
the threshold for breakup into the neutron core continuum. 
}
\label{fig:kCotDeltaOne}
\end{center}
\end{figure*}
 
\begin{figure*}[bthp]
\begin{center}
\includegraphics [scale=0.46] {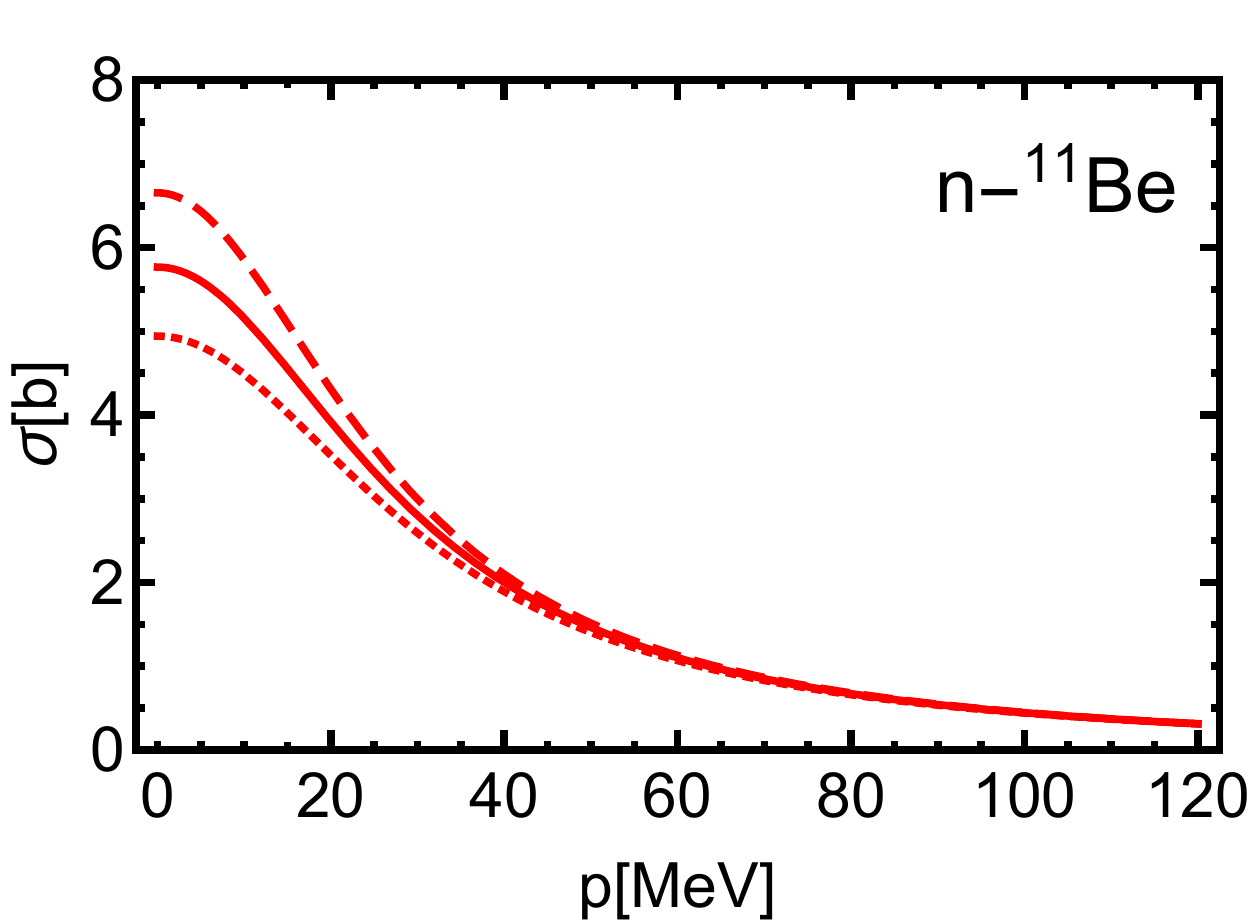}
\includegraphics [scale=0.46] {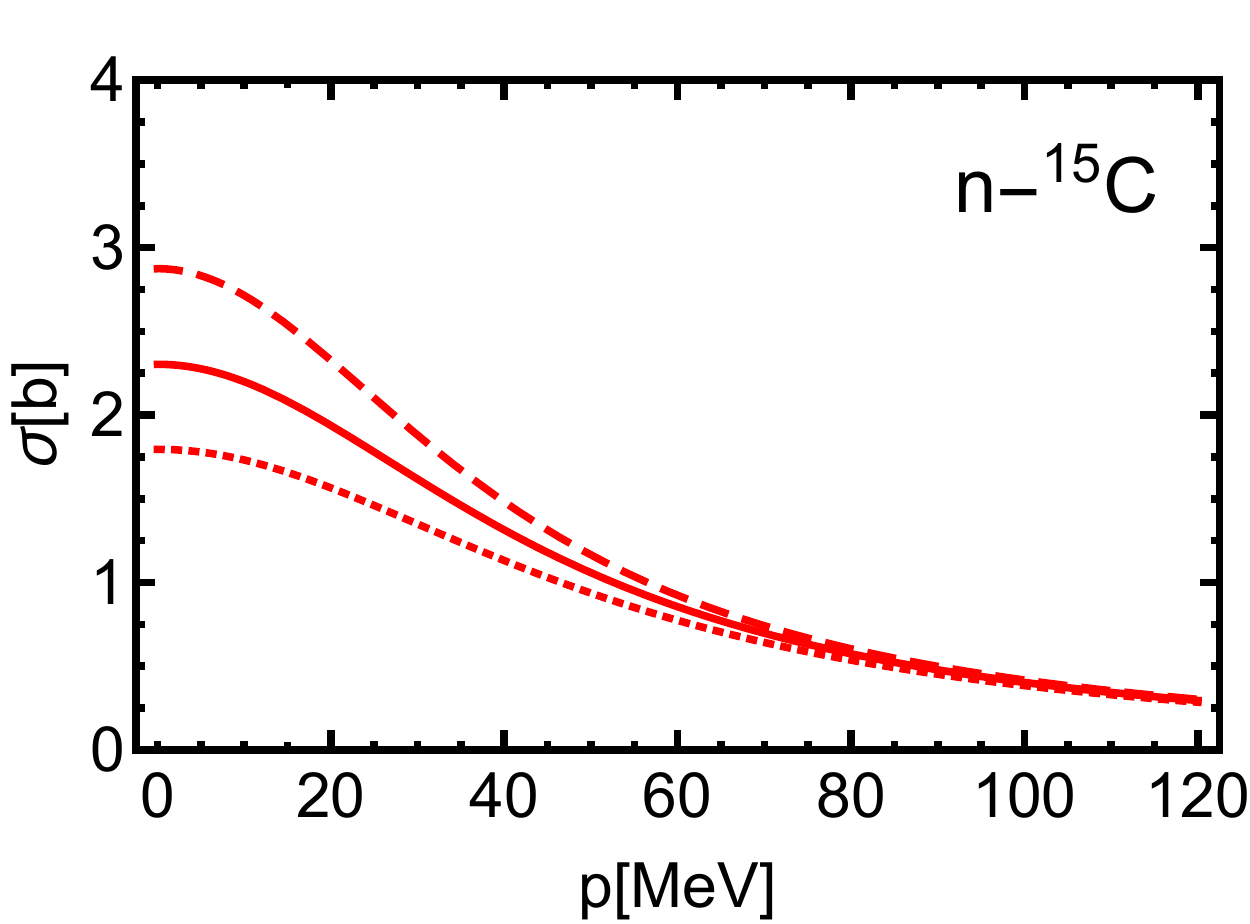}
\includegraphics [scale=0.46] {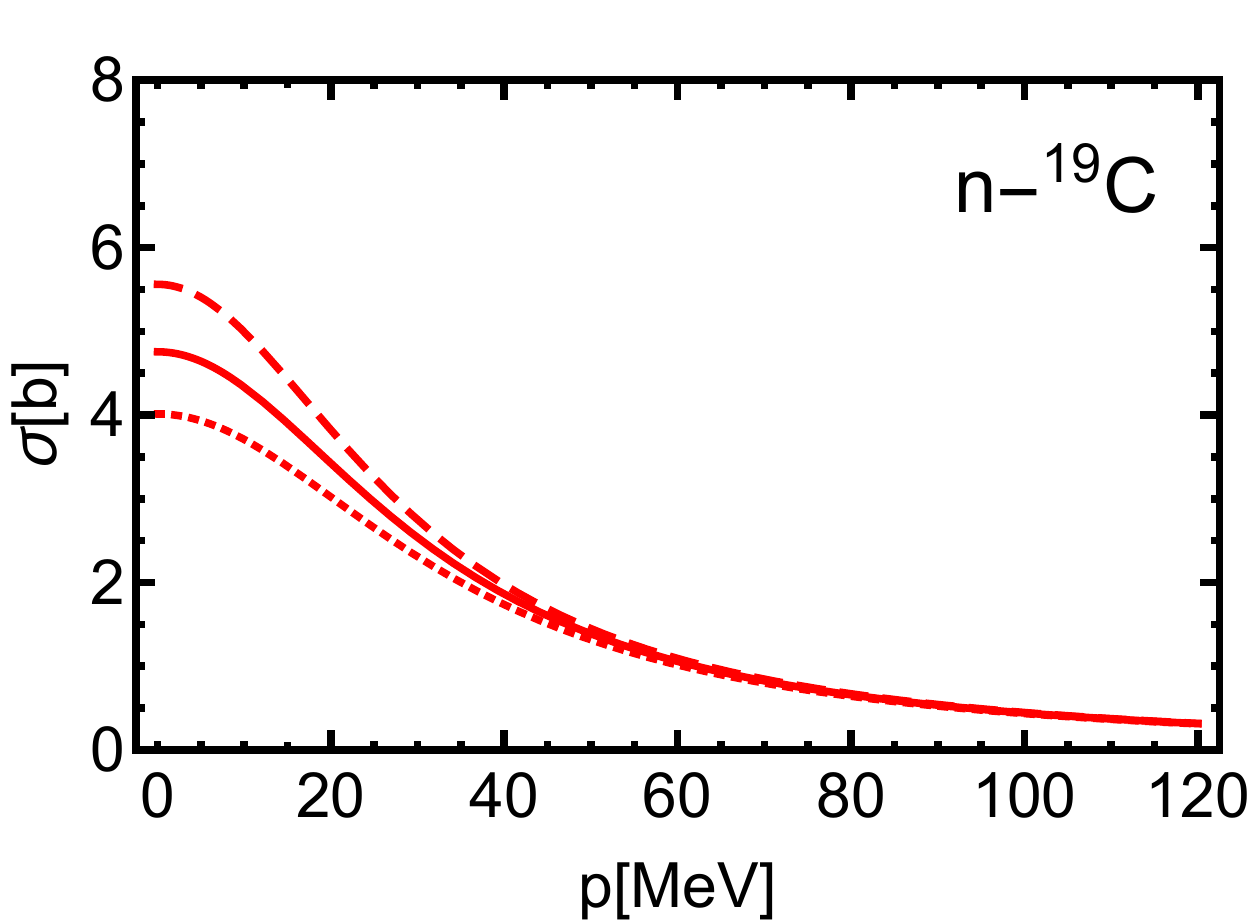}\\
\caption{The total $s$-wave cross sections for the  scattering of neutron and one-neutron halo nuclei in the $J=1$ channel: $n\text{-}^{11}\rm Be$ (left), $n\text{-}^{15}\rm C$ (middle), and $n\text{-}^{19}\rm C$ (right). 
The notations of the solid, dashed and dotted curves are the same as those in Fig.~\ref{fig:kCotDeltaOne}. 
}
\label{fig:CrossSecOne}
\end{center}
\end{figure*}

\subsection{Amplitudes and total cross sections in the $J=1$ channel}

The on-shell amplitudes for scattering of neutrons from the neutron halo nuclei, $T$, are related to the scattering phase shifts through the relation
\begin{equation}
T_{\sigma\sigma}^{\,l}(p,p,E)=\frac{2\pi}{\mu_{n\sigma}}\frac{1}{p\cot\delta_{\,l}(p)-ip}, \label{eq:pcot}
\end{equation}
where the reduced mass $\mu_{n\sigma}$ is defined in Eq.~\eqref{Eq:nsigmared}.
The 
differential cross section for a given spin state in terms of the phase shifts can be written as~\cite{Goldberger:1964}
\begin{equation}
\frac{d\sigma}{d\Omega}=\sum_l\bigg|\frac{2l+1}{p\cot\delta_{\,l}(p)-ip}P_l(\cos\theta)\bigg|^2.
\end{equation}
Then the 
total cross section is
\begin{equation}
\sigma(p,p,E)=\sum_l\frac{(2l+1)\mu_{n\sigma}^2}{\pi}\big|T_{\sigma\sigma}^{\,l}(p,p,E)|^2.
\end{equation}

In the $J=1$ channel, no three-body bound states in the $nn\rm ^{10}Be$, $nn\rm ^{14}C$ and $nn\rm ^{18}C$ systems exist.
The equations for the amplitudes of $n$-$^{11}\rm Be$, $n$-$^{15}\rm C$ and $n$-$^{19}\rm C$ scattering have a unique solution as $\Lambda\to \infty$.
We take the values in Table~\ref{Tab:InputA} as input for the the $nc$ scattering length $a_{\sigma}$ and vary it by $\pm0.5$~fm to show the dependence of the results on this input. 
The results of $p\cot\delta_0(p)$ for the $s$-wave scattering defined in Eq.~\eqref{eq:pcot} are shown in Fig.~\ref{fig:kCotDeltaOne}.
One can see that $p\cot\delta_0(p)$ (and also the phase shift $\delta_0(p)$) is real below the breakup threshold of the one-neutron halo nucleus and develops a small imaginary part above since the neutron-core continuum channel is open. The real part of $p\cot\delta_0(p)$ is
essentially constant over the full range of $p$ considered,
resulting in a very small $n\sigma$ effective range.
Finally, the results depend only weakly on $a_\sigma$.

The total $s$-wave scattering cross sections $\sigma$ are shown in Fig.~\ref{fig:CrossSecOne}.
One can see that there is a bump at threshold in each of the total cross sections of the
$n$-$^{11}\rm Be$, $n$-$^{15}\rm C$ and $n$-$^{19}\rm C$ scattering processes, and the total cross sections have a small change in the ranges of $a_{\sigma}$ we considered. Our numerical results suggest that 
the total cross sections at threshold are of the order of a few barns for the $n$-$^{11}\rm Be$, $n$-$^{15}\rm C$ and $n$-$^{19}\rm C$ scattering.

\subsection{Amplitudes and total cross sections in the $J=0$ channel}

In the $J=0$ channel, the amplitudes of the $n$-$^{11}\rm Be$, $n$-$^{15}\rm C$ and $n$-$^{19}\rm C$ scattering
do not have a unique solution as $\Lambda\to \infty$ if the $D_0$ term in Eq.~\eqref{eq:C0} is neglected. 
For finite $\Lambda$, they display a strong cutoff
dependence.
The three-body counterterm $D_0$ is required for renormalization,
as discussed in 
Sec.~\ref{sec:CounterTerm}, and the value for each of the $\rm ^{12}Be$, $\rm ^{16}C$ and $\rm ^{20}C$ systems is fixed through the two-neutron separation energies listed in Table~\ref{Tab:InputB}. 

\begin{figure*}[ht]
\begin{center}
\includegraphics [scale=0.45] {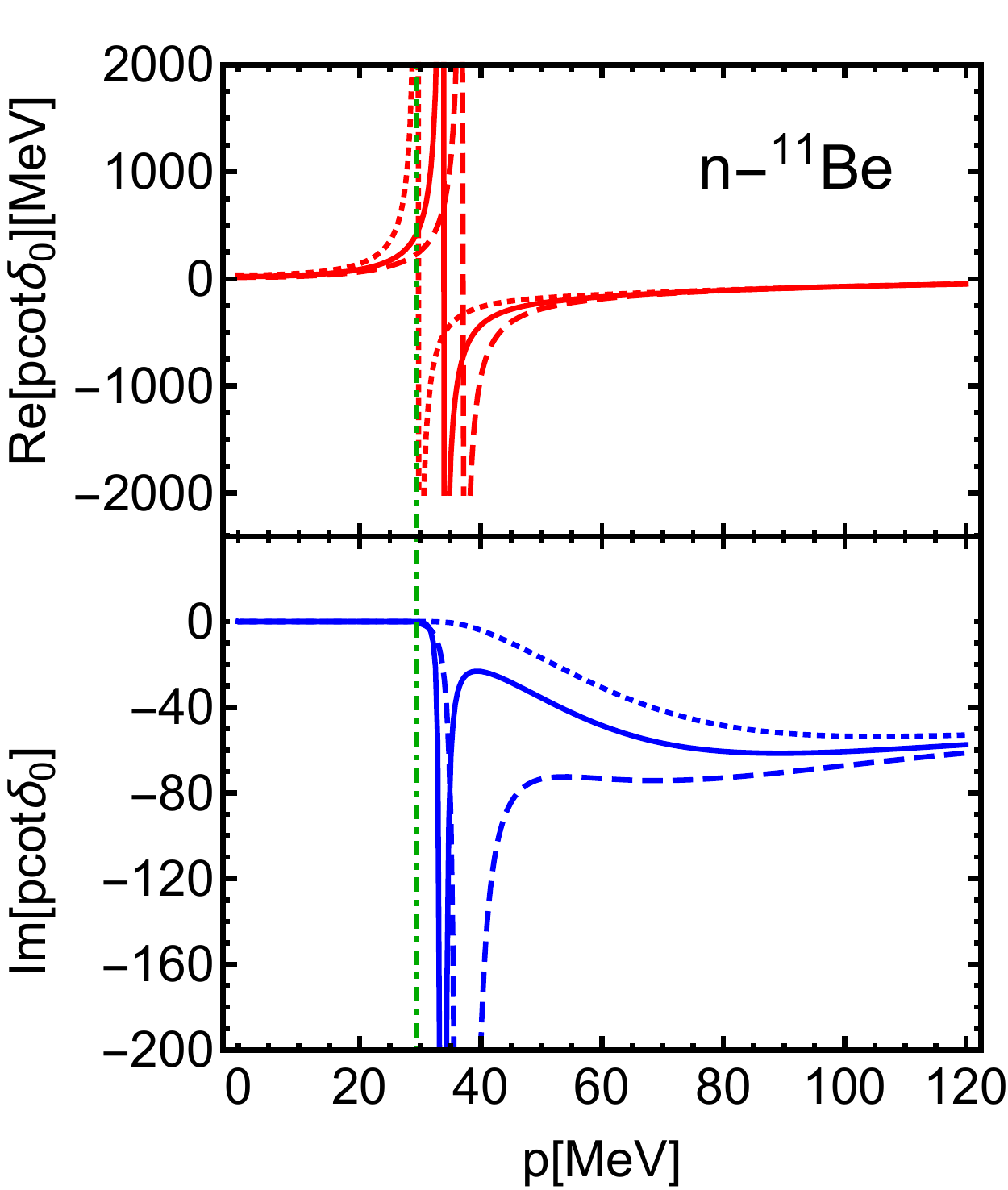}
\includegraphics [scale=0.45] {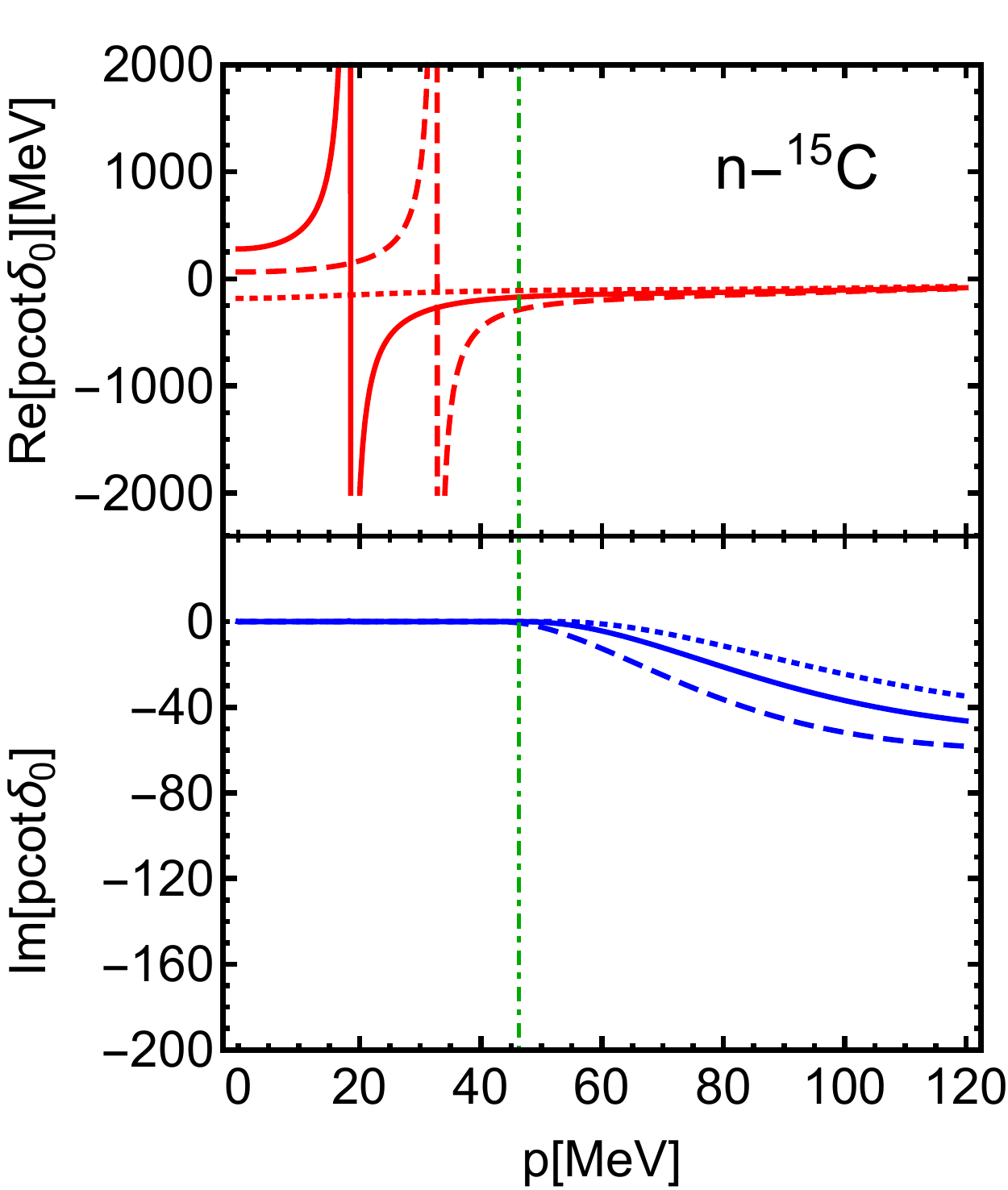}
\includegraphics [scale=0.45] {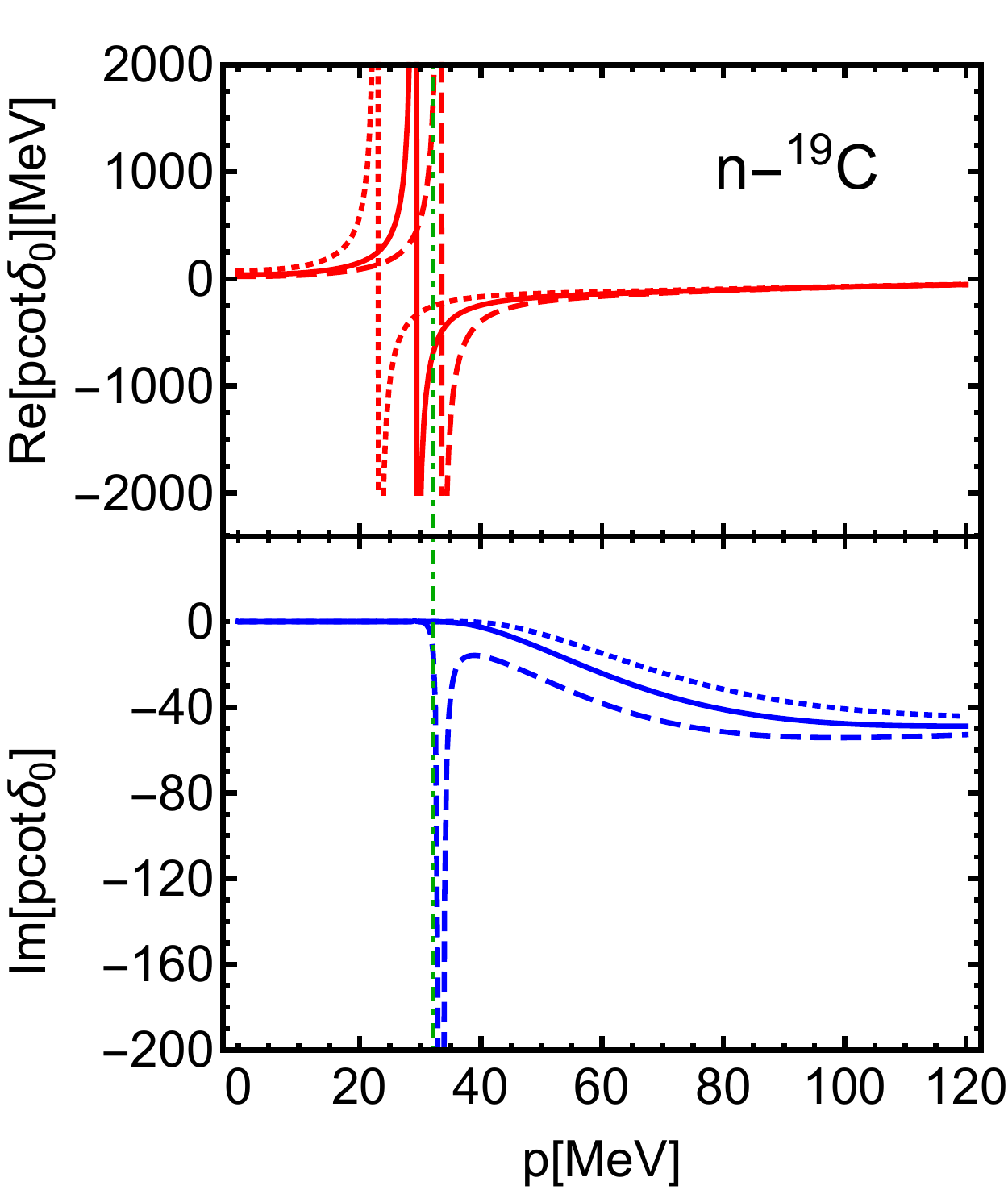}
\caption{Real and imaginary parts of $p\cot\delta_0(p)$ for the $s$-wave scattering of neutron and one-neutron halo nuclei in the $J=0$ channel. The left, middle and right panels correspond to the $n\text{-}^{11}\rm Be$, $n\text{-}^{15}\rm C$ and $n\text{-}^{19}\rm C$ scattering processes, respectively.
The three-body forces are fixed through the two-neutron separation energies of $\rm ^{12}Be$, $\rm ^{16}C$ and $\rm ^{20}C$ in Table~\ref{Tab:InputB}. 
The notations of the solid, dashed, dotted and dot-dashed lines are the same as those in Fig.~\ref{fig:kCotDeltaOne}.
}
\label{fig:kCotDeltaZero}
\end{center}
\end{figure*}

The results for the $s$-wave scattering $p\cot\delta_0 (p)$ in the $J=0$ case are shown in Fig.~\ref{fig:kCotDeltaZero}, where the solid lines represent the results using the $a_\sigma$ values in Table~\ref{Tab:InputA} as inputs and the dashed (dotted) lines correspond to those obtained by increasing (decreasing) $a_\sigma$ by 0.5~fm.

\begin{figure*}[ht]
\begin{center}
\includegraphics [scale=0.46] {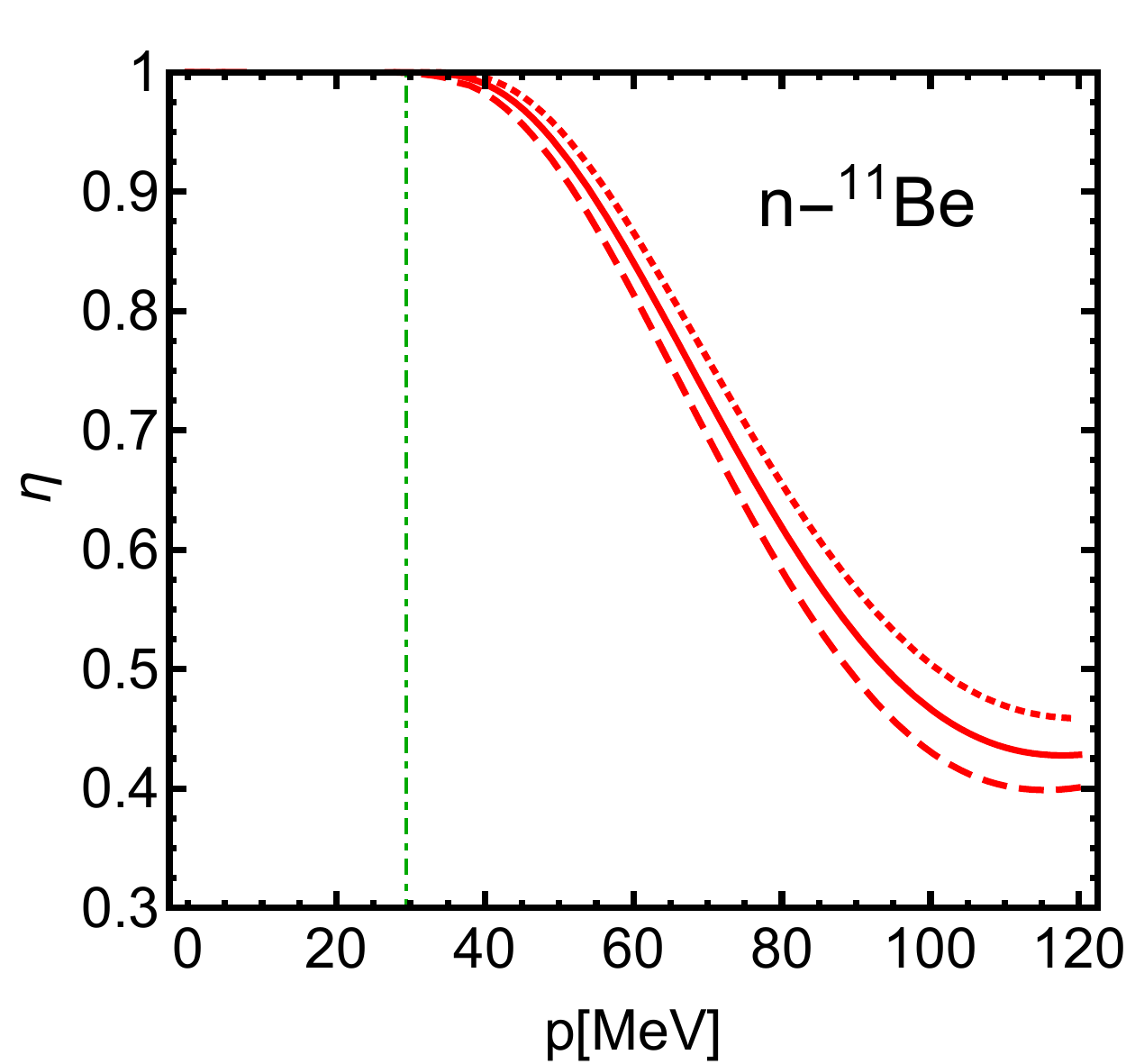}
\includegraphics [scale=0.46] {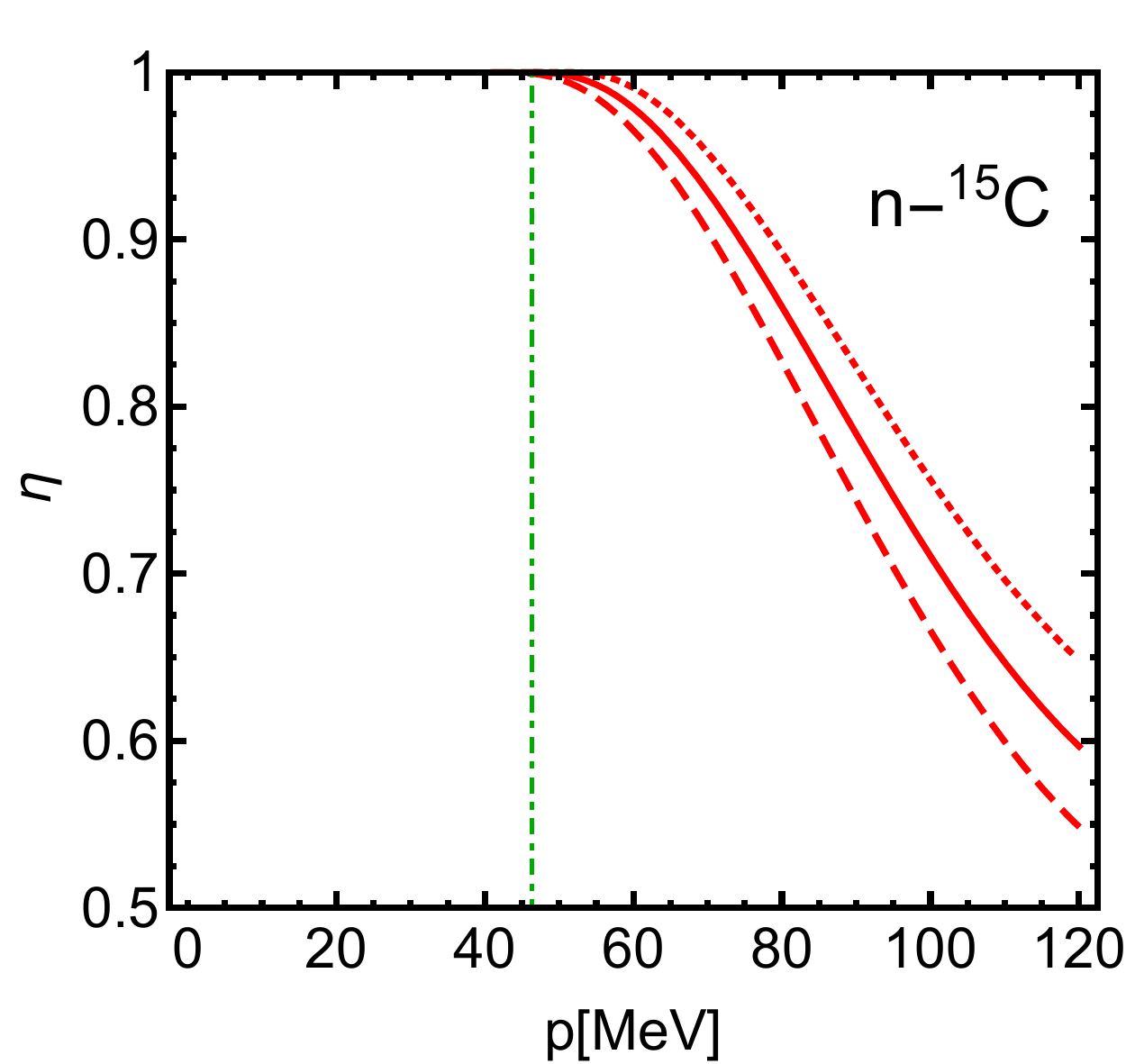}
\includegraphics [scale=0.46] {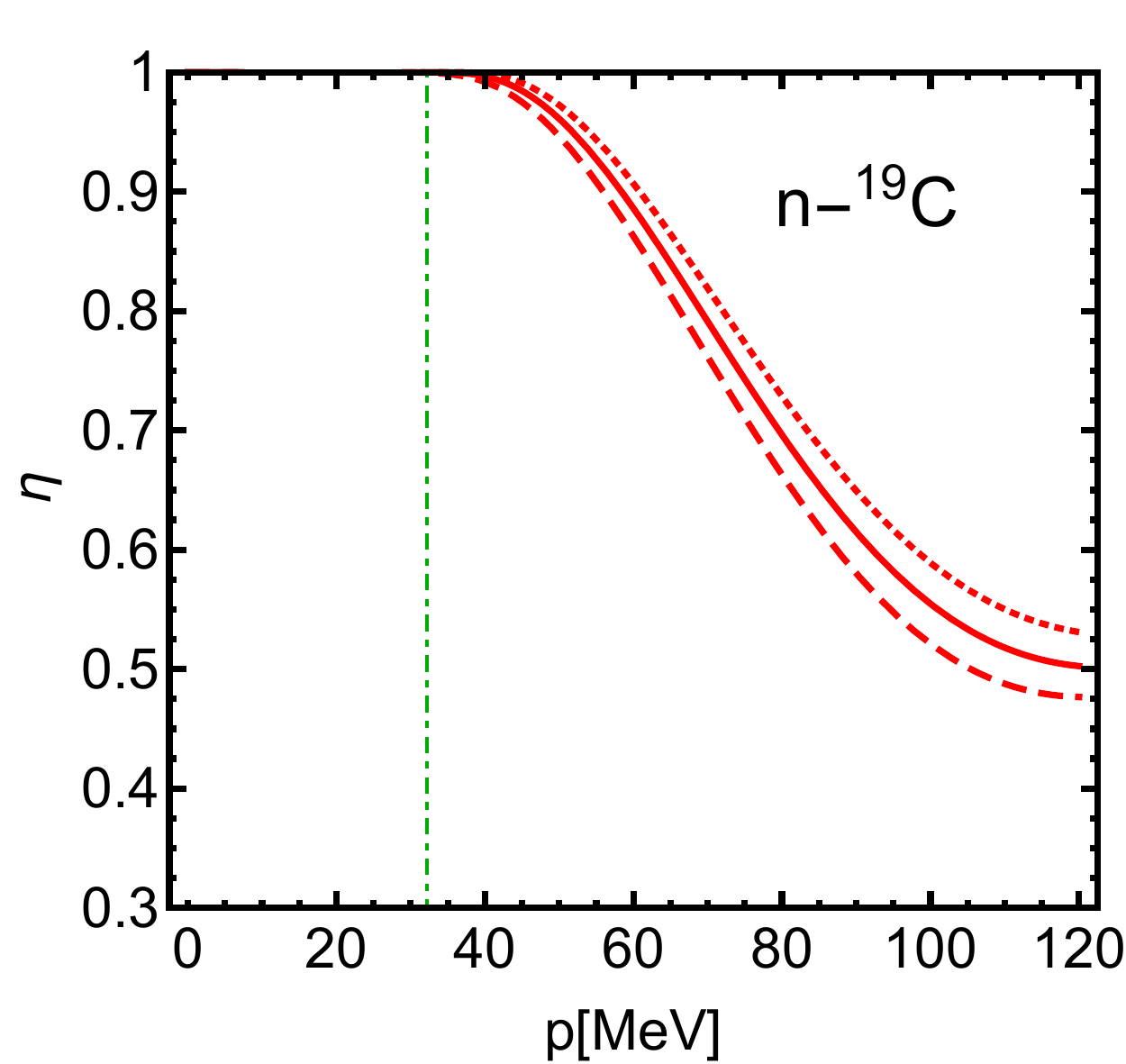}\\
\caption{Inelasticity factor for $s$-wave scatterings of the neutron and one-neutron halo nuclei in the $J=0$ channel: $n\text{-}^{11}\rm Be$ (left), $n\text{-}^{15}\rm C$ (middle), and $n\text{-}^{19}\rm C$ (right).
The notations of the solid, dashed, dotted and dot-dashed lines are the same as those in Fig.~\ref{fig:kCotDeltaOne}.}
\label{fig:eta}
\end{center}
\end{figure*}

\begin{figure*}[ht]
\begin{center}
\includegraphics [scale=0.46] {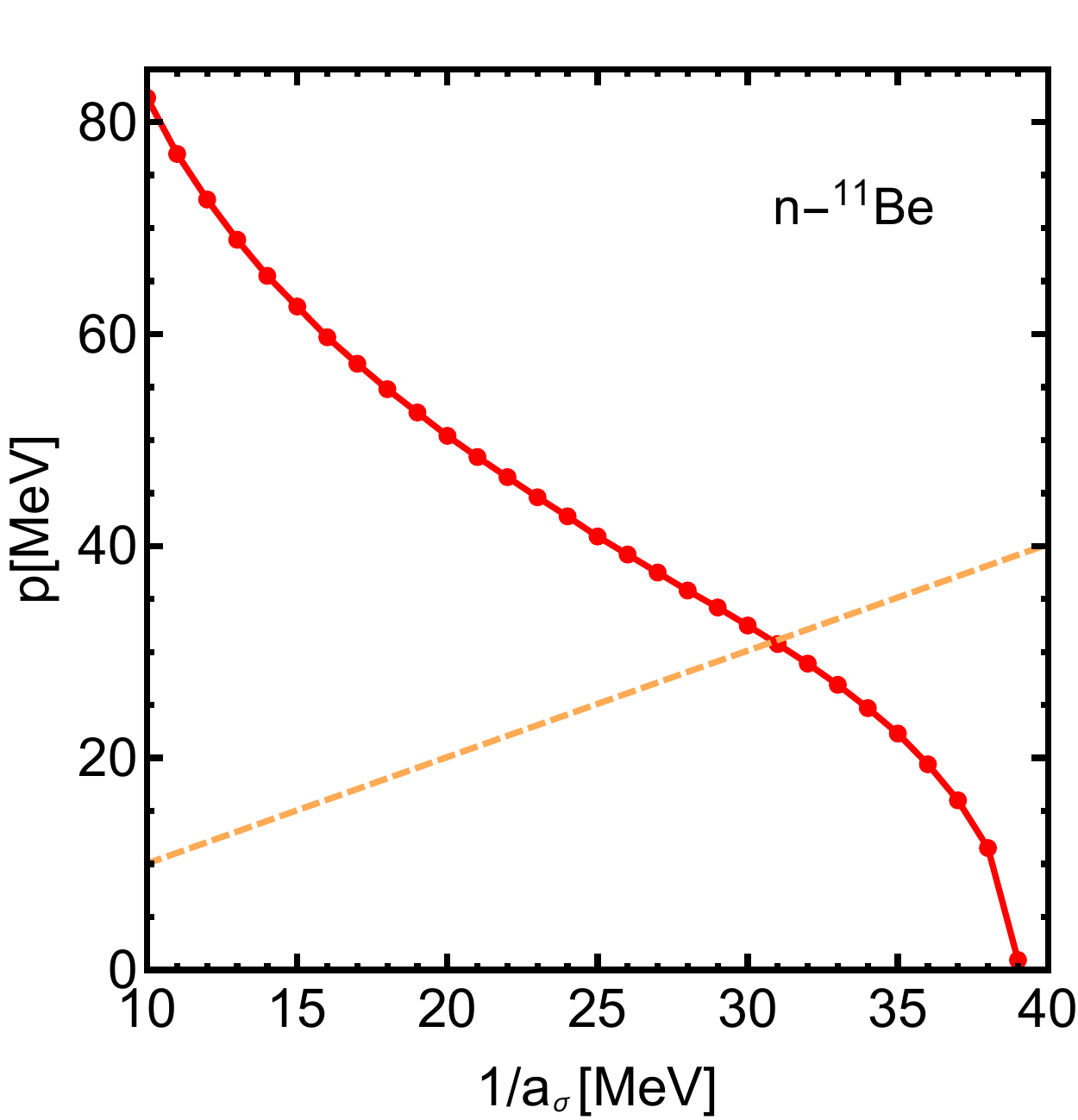}
\includegraphics [scale=0.46] {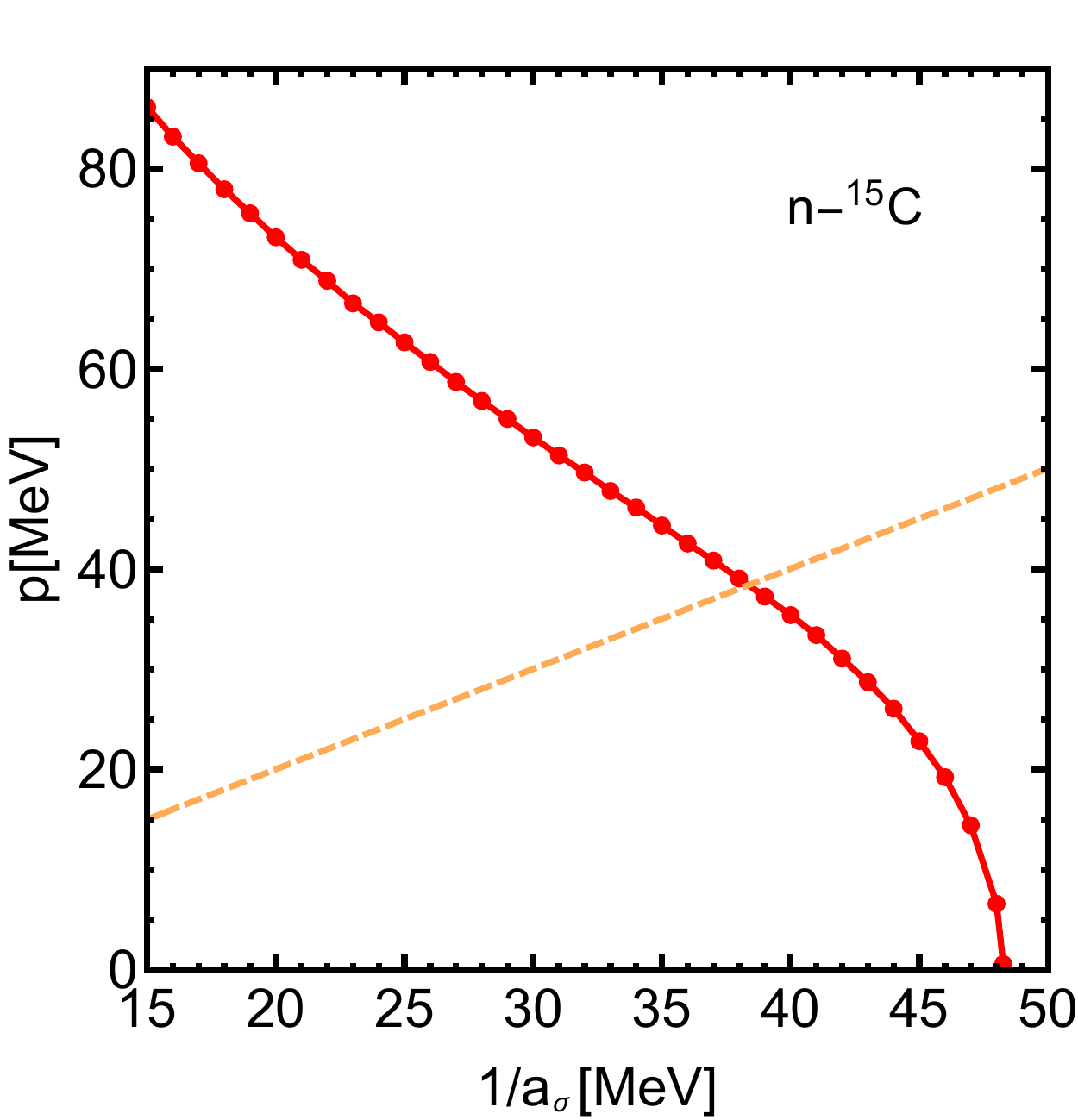}
\includegraphics [scale=0.46] {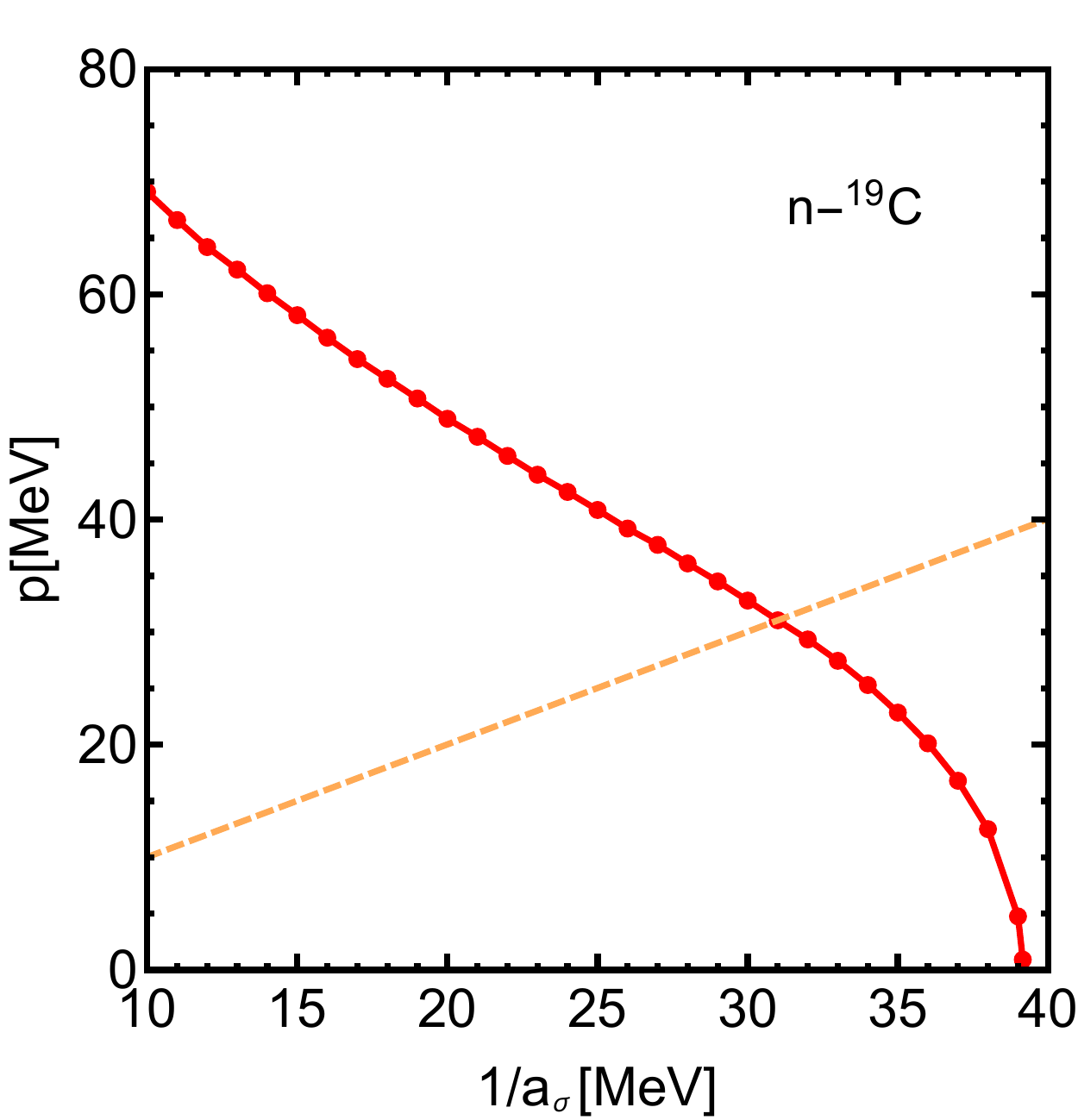}\\
\caption{Location of the pole in $p\cot\delta_0^R(p)$ as a function of $1/a_{\sigma}$: $n\text{-}^{11}\rm Be$ (left), $n\text{-}^{15}\rm C$ (middle), and $n\text{-}^{19}\rm C$ (right). The dashed line represents 
the threshold for breakup into the neutron core continuum. 
}
\label{fig:trajectory}
\end{center}
\end{figure*}

The values of $p\cot\delta_0(p)$ vare real below the threshold for breakup of the one-neutron halo nuclei in the scattering process. They become complex above since the breakup channel is open, and compared to the $J=1$ channel the imaginary parts of $p\cot\delta_0(p)$ are much larger.

\begin{figure*}[ht]
\begin{center}
\includegraphics [scale=0.46] {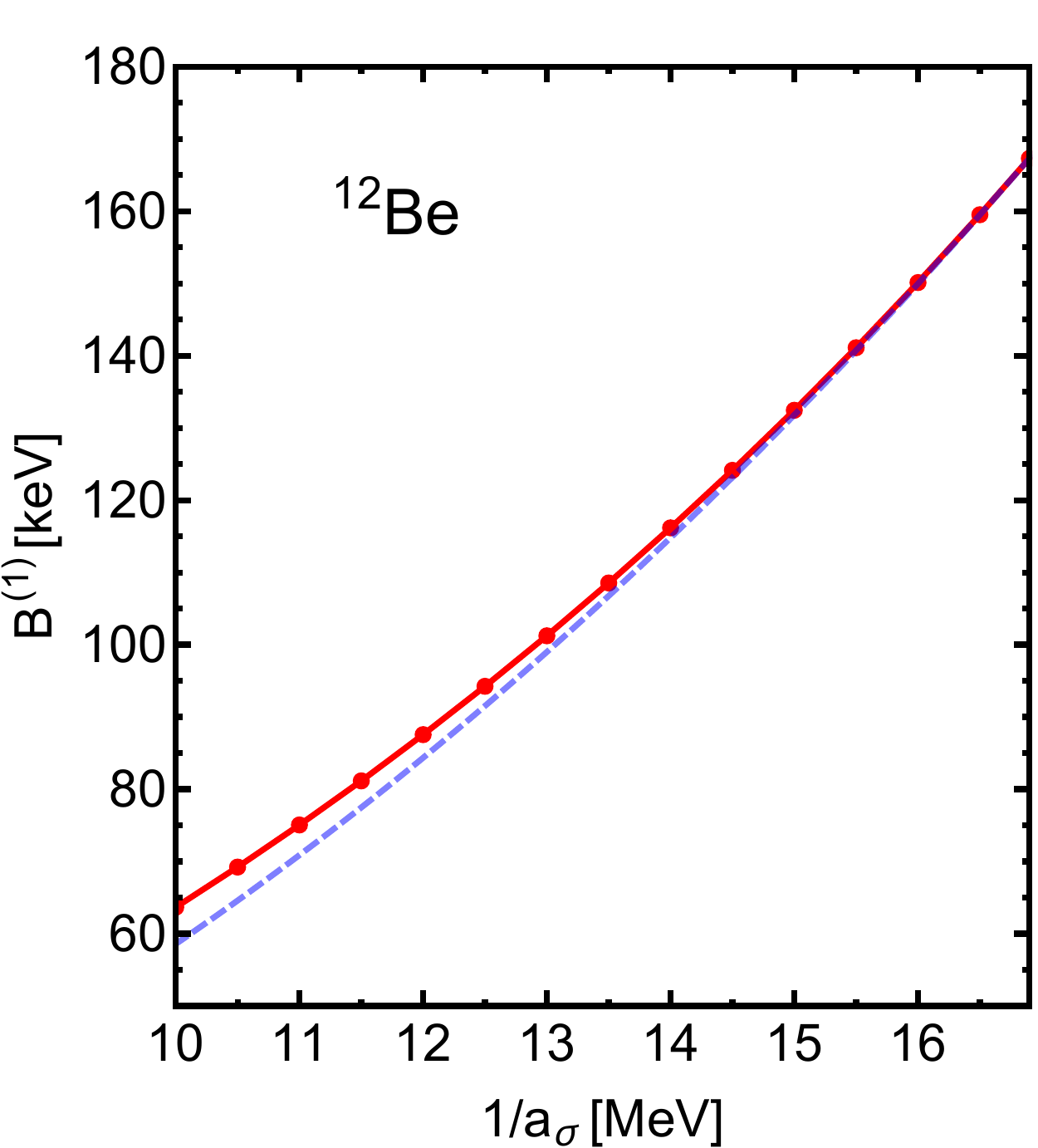}
\includegraphics [scale=0.46] {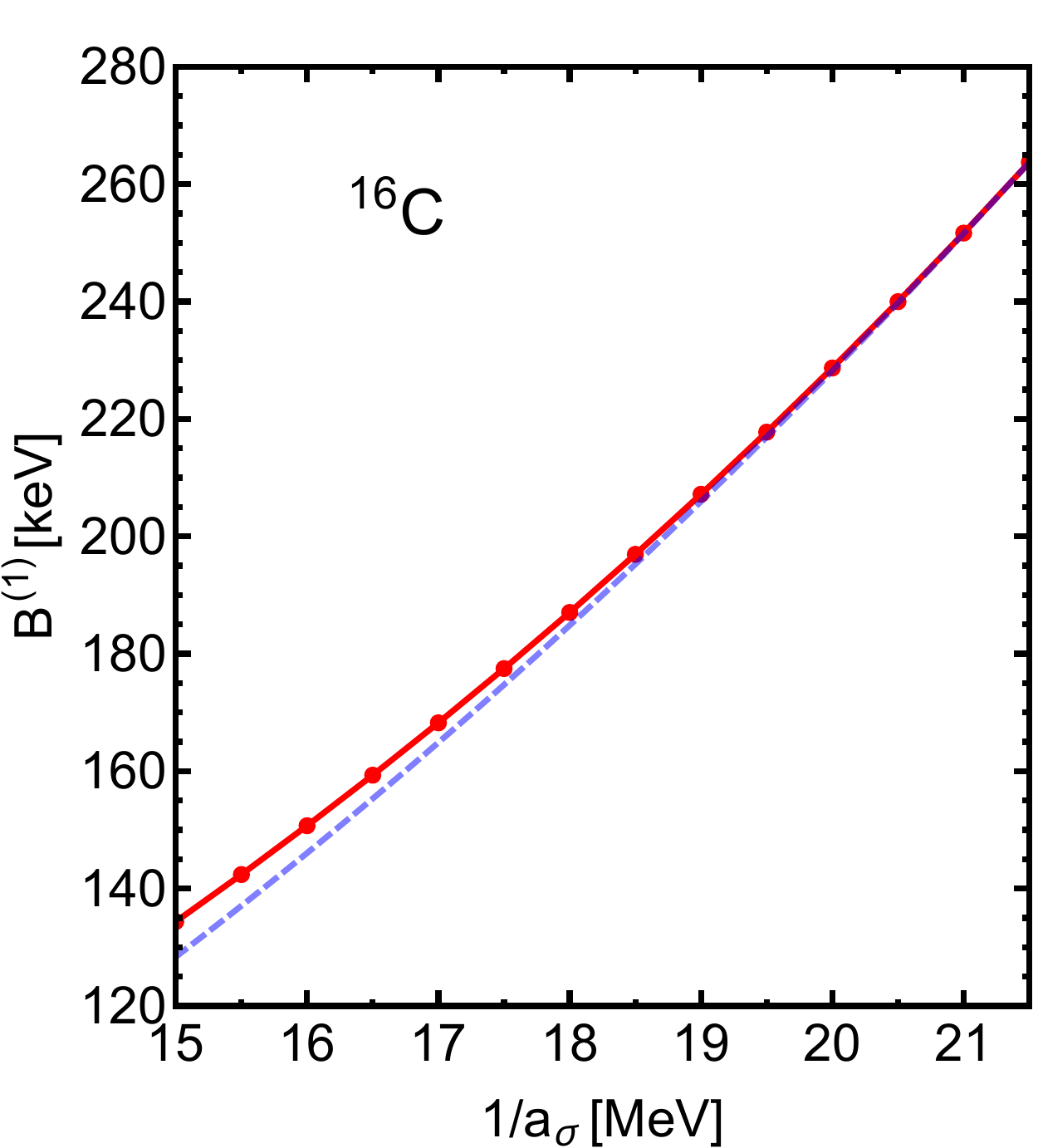}
\includegraphics [scale=0.46] {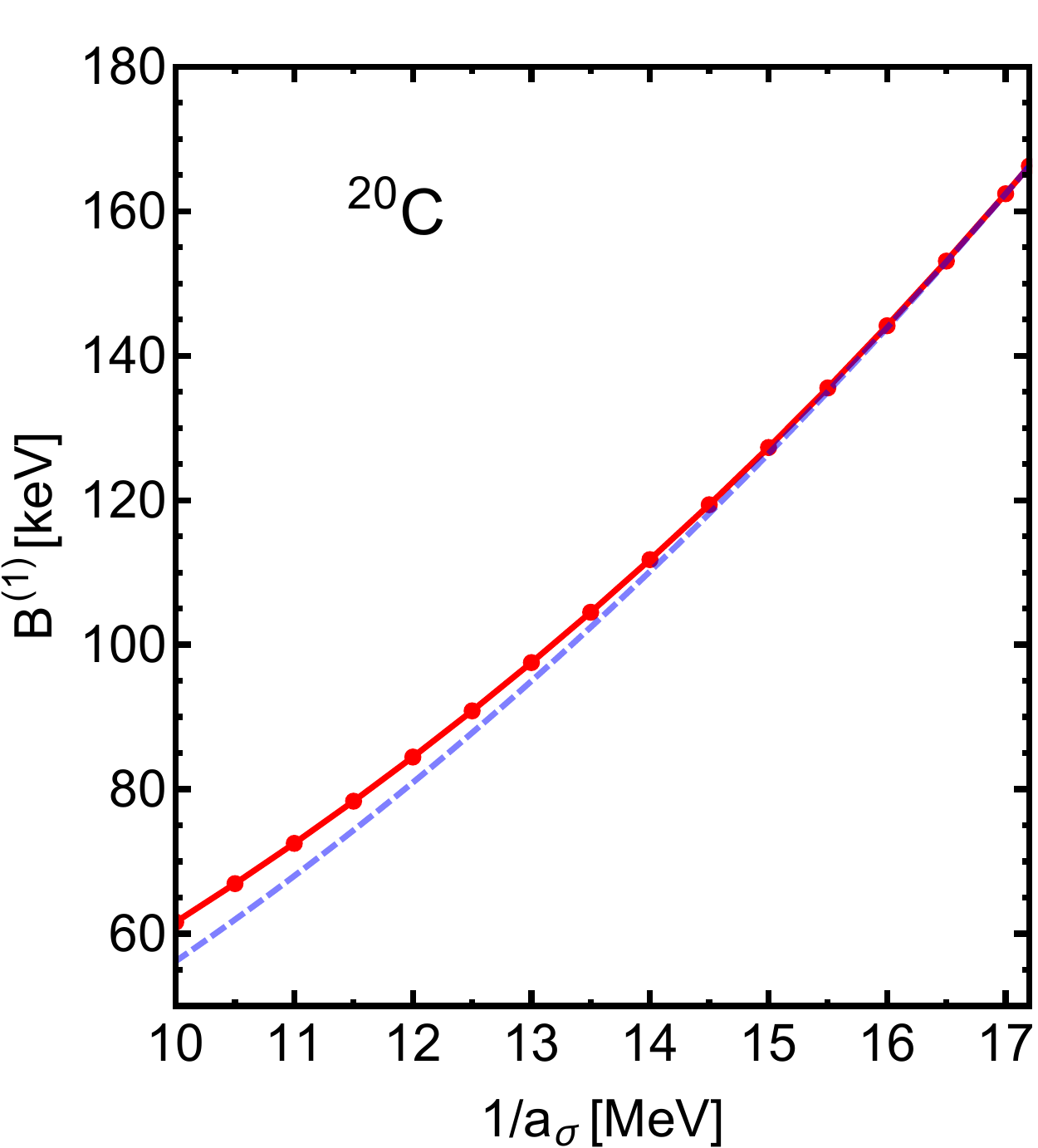}
\caption{Binding energy of the excited state as a function of $1/a_{\sigma}$: $^{12}\rm Be$ (left), $^{16}\rm C$ (middle), and $^{20}\rm C$ (right). The dashed line represents the scattering threshold which is given by $B^{(1)}=B_{\sigma}$.}
\label{fig:excitedstate}
\end{center}
\end{figure*}

\begin{figure*}[ht]
\begin{center}
\includegraphics [scale=0.46] {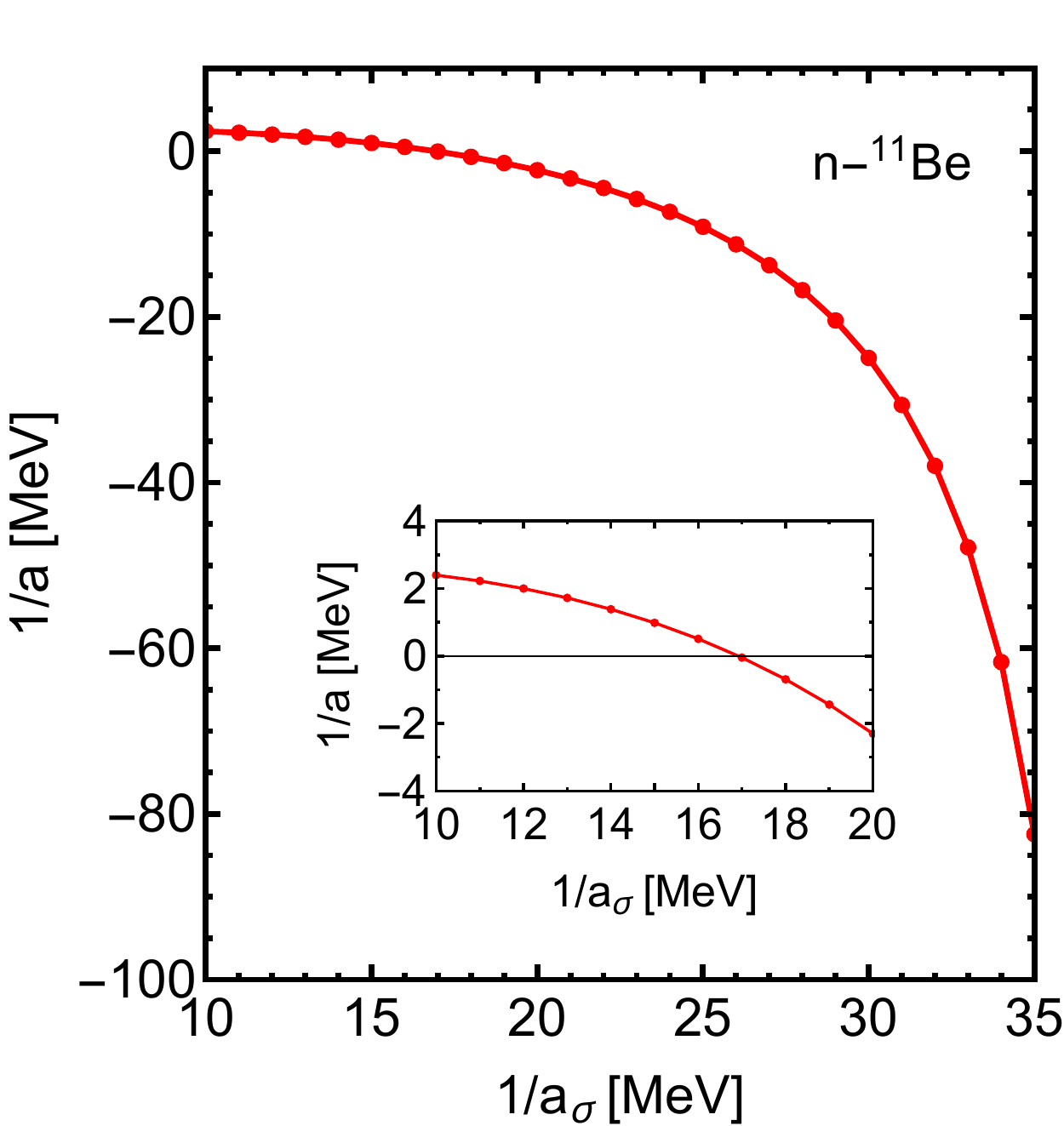}
\includegraphics [scale=0.46] {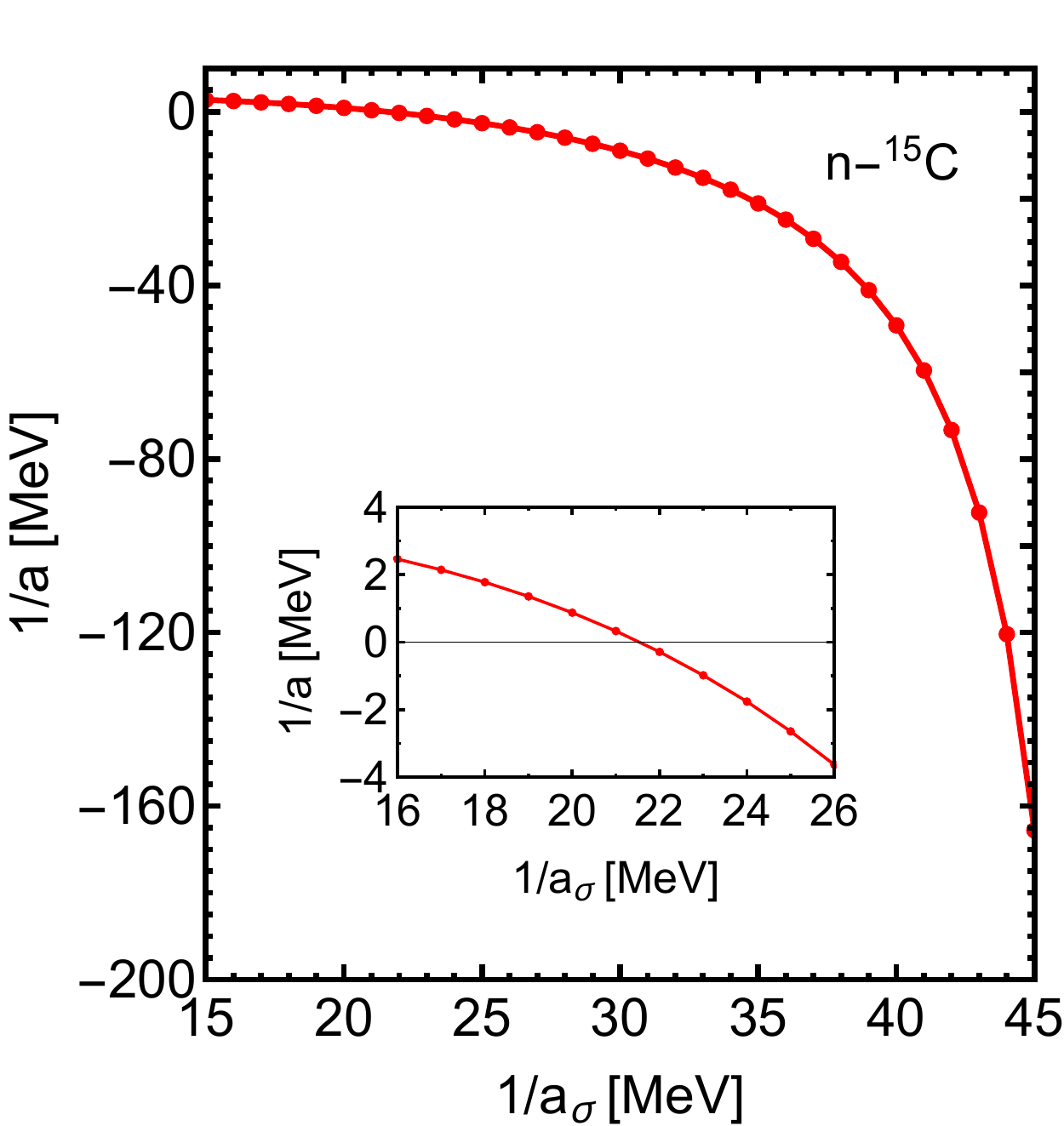}
\includegraphics [scale=0.46] {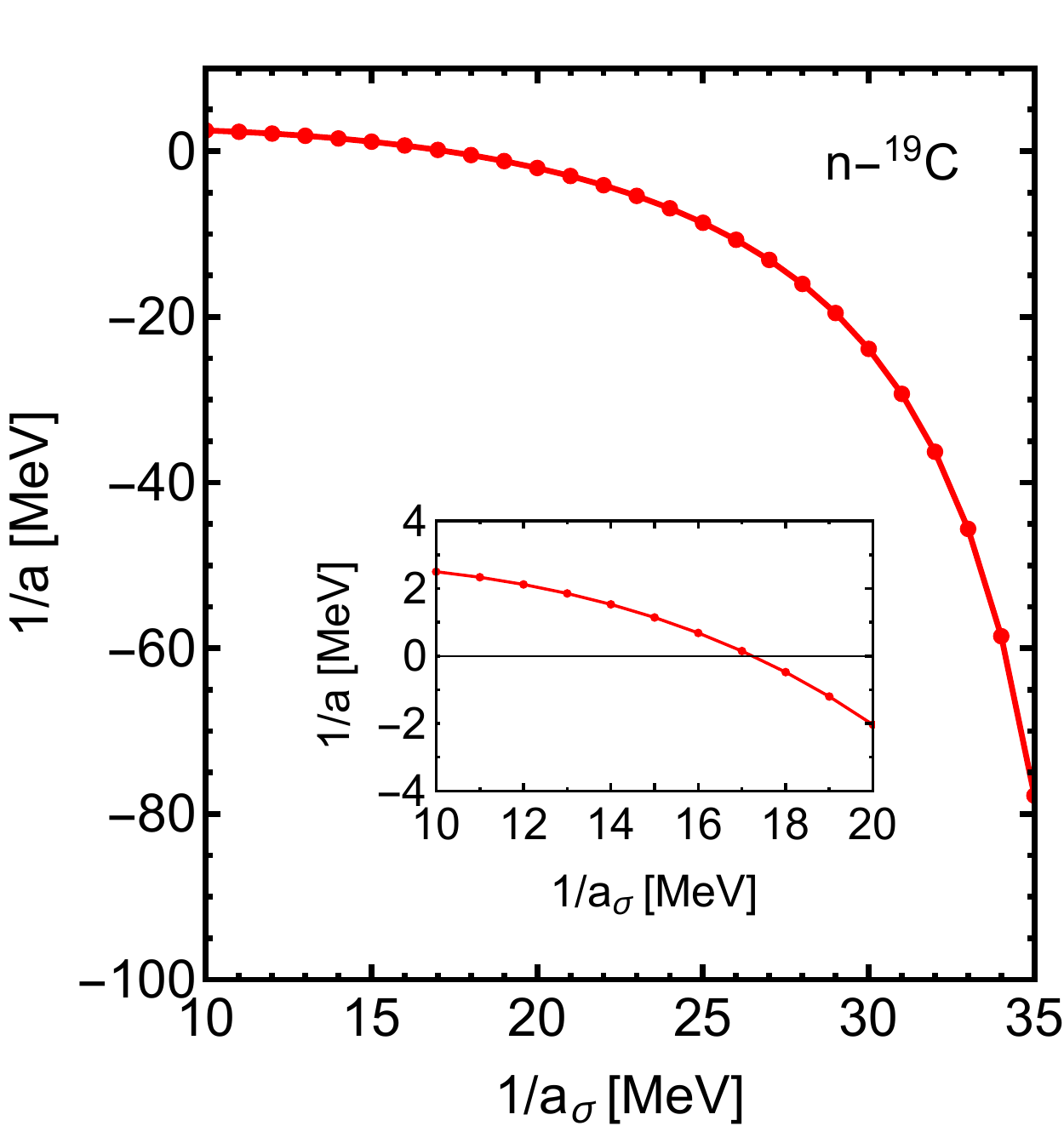}
\caption{Inverse of the scattering length $1/a$ of the $s$-wave scatterings of the neutron and one-neutron halo nuclei in the $J=0$ channel: $n\text{-}^{11}\rm Be$ (left), $n\text{-}^{15}\rm C$ (middle), and $n\text{-}^{19}\rm C$ (right). The inset shows the transition of $1/a$ from positive to negative values in more detail.}
\label{fig:ScatteringLength}
\end{center}
\end{figure*}

Alternatively, the on-shell amplitudes $T$ can be related to real-valued scattering phase shifts $\delta_l^R(p)$ through the relation 
\begin{equation}
T_{\sigma\sigma}^{\,l}(p,p,E)=\frac{2\pi}{\mu_{n\sigma}}\frac{\eta e^{2i\delta_l^R(p)}-1}{2ip}, \label{eq:pcotreal}
\end{equation}
with $\eta$ is the inelasticity factor. With this definition $\eta=1$ below the breakup threshold and $0\le \eta<1$ above. Thus a large inelasticity correponds to small values of $\eta$. 
Our results for $\eta$ are shown in Fig.~\ref{fig:eta}. The inelasticity grows rapidly 
when the momentum $p$ is increased beyond the breakup threshold.

The most interesting feature of our results is that for all the considered processes, there is a pole in $p\cot\delta_0^R(p)$, and thus a zero of the $T$-matrix when this pole is below the breakup threshold. The $T$-matrix
almost vanishes at the pole position when the inelasticity is small.
The pole position is sensitive to $a_\sigma$.
It can be seen from, {\it e.g.}, the results for $n$-$^{15}$C scattering (the middle plot in Fig.~\ref{fig:kCotDeltaZero}), that the pole disappears when $a_\sigma$ decreases from 4.27~fm in Table~\ref{Tab:InputA} to 3.77~fm.
Increasing $a_\sigma$ pushes the pole to larger values of $p$, as can be seen from Fig.~\ref{fig:trajectory} which shows the location of the pole in $p\cot\delta_0^R(p)$ as a function of $1/a_{\sigma}$. 
Our results are in qualitative agreement with the findings of Refs.~\cite{Yamashita:2008sg,Shalchi:2016psb,Deltuva:2017zvk} for the  case of $n$-$^{19}$C scattering.

The appearance of this pole indicates the presence of an excited virtual Efimov state close to the scattering threshold.  
As $a_\sigma$ is increased the virtual excited state turns into a real excited state that becomes part of the bound state spectrum (this can be understood from, {\it e.g.}, Fig.~23 of Ref.~\cite{Braaten:2004rn}). 
Indeed, an excited bound state is found as $a_\sigma$ is increased. We calculate the value of the excited state energy as a function of $1/a_{\sigma}$, with the three-body force fixed through the two-neutron separation energies of $\rm ^{12}Be$, $\rm ^{16}C$ and $\rm ^{20}C$ in Table~\ref{Tab:InputB}. The dependence of the binding energy of the excited state as a function of $1/a_\sigma$ is plotted in Fig.~\ref{fig:excitedstate}. 
We find that for $\rm ^{12}Be$ the first excited state appears when $a_{\sigma}$ is increased beyond the physical value, {\it viz.} $1/a_{\sigma}\lesssim 16.9$~MeV, while
for $^{16}\rm C$ and $\rm ^{20}C$ the excited state appears for $1/a_{\sigma}\lesssim 21.5$~MeV and $1/a_{\sigma}\lesssim 17.2$~MeV, respectively. 
The appearance of the excited three-body bound state is also signaled by a sign change of the $s$-wave scattering length for scattering of neutrons from the neutron halo, $a$, defined as $p\cot\delta_0(p)\overset{p \to 0}{=}-1/a$: $a$ is negative if there is only the three-body ground state of the halo nucleus and turns positive when the virtual three-body state becomes a bound excited state. 
The results for the $s$-wave $n$-$^{11}\rm Be$, $n$-$^{15}\rm C$ and $n$-$^{19}\rm C$ scattering length $a$ are shown in Fig.~\ref{fig:ScatteringLength}. 
Thus the experimental observation of such a pole in $p\cot\delta_0^R(p)$ could serve as a confirmation 
of Efimov physics in halo nuclei, even if no bound excited state is present. Up to higher order corrections, the energy of the virtual state is completely determined by the ground state energies of the one- and two-neutron halo nuclei under consideration and the neutron-neutron scattering length.

\begin{figure*}[tbh]
\begin{center}
\includegraphics [scale=0.46] {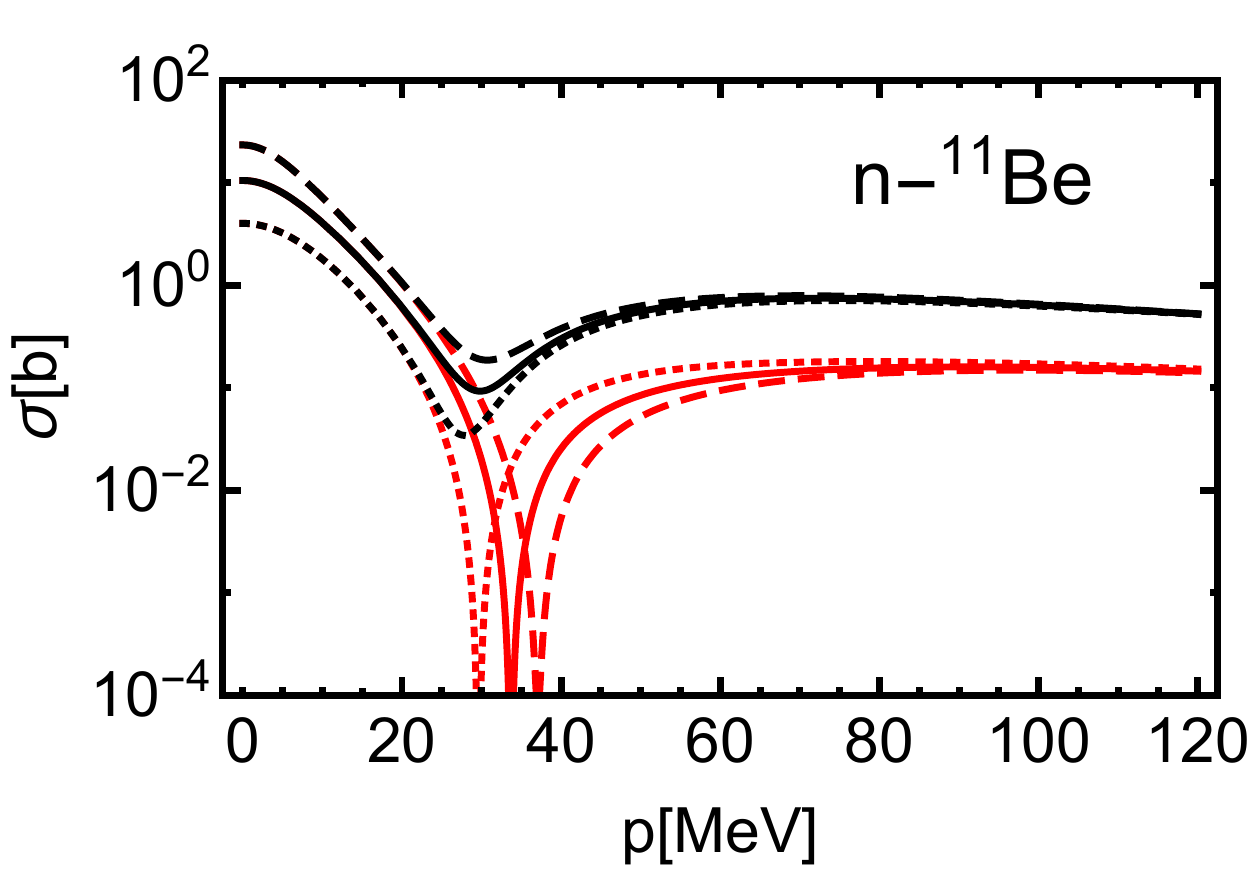}
\includegraphics [scale=0.46] {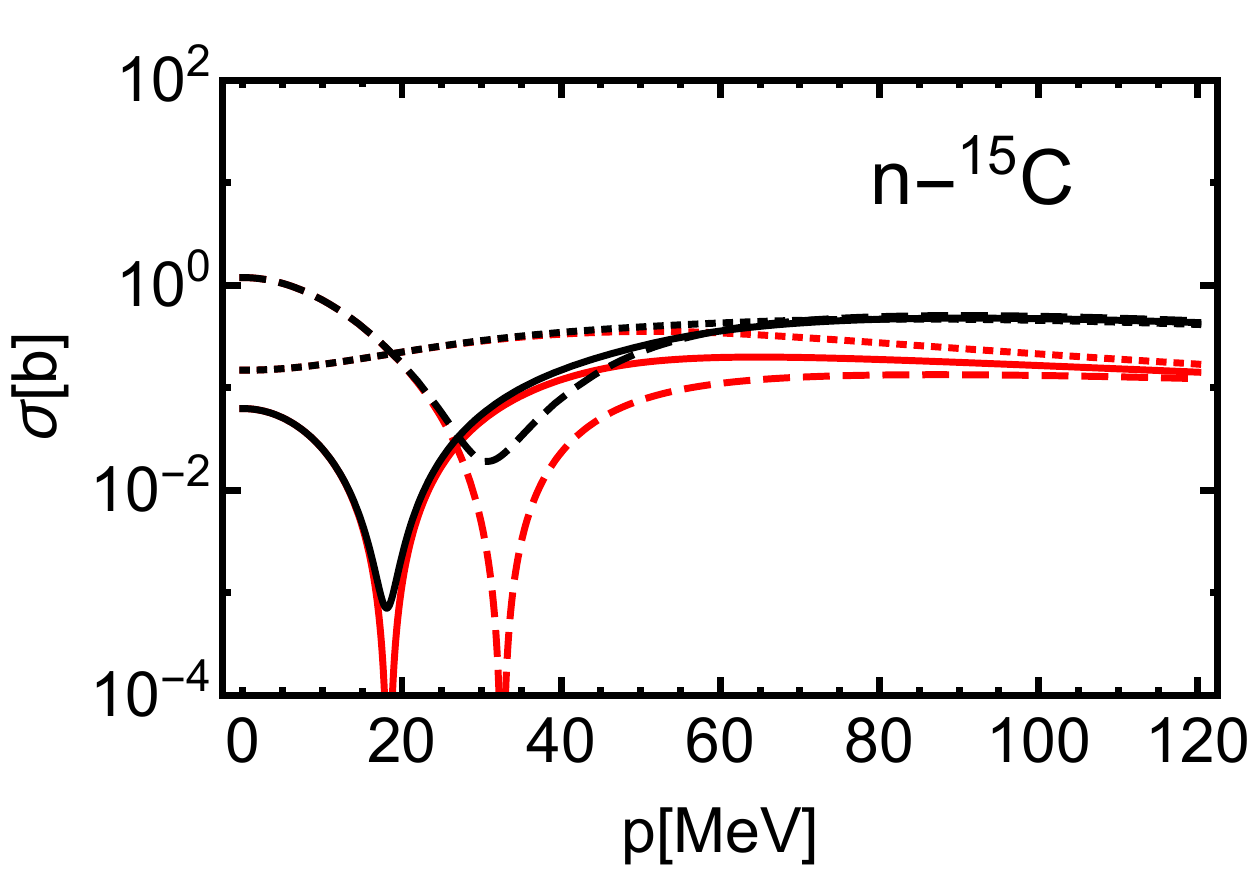}
\includegraphics [scale=0.46] {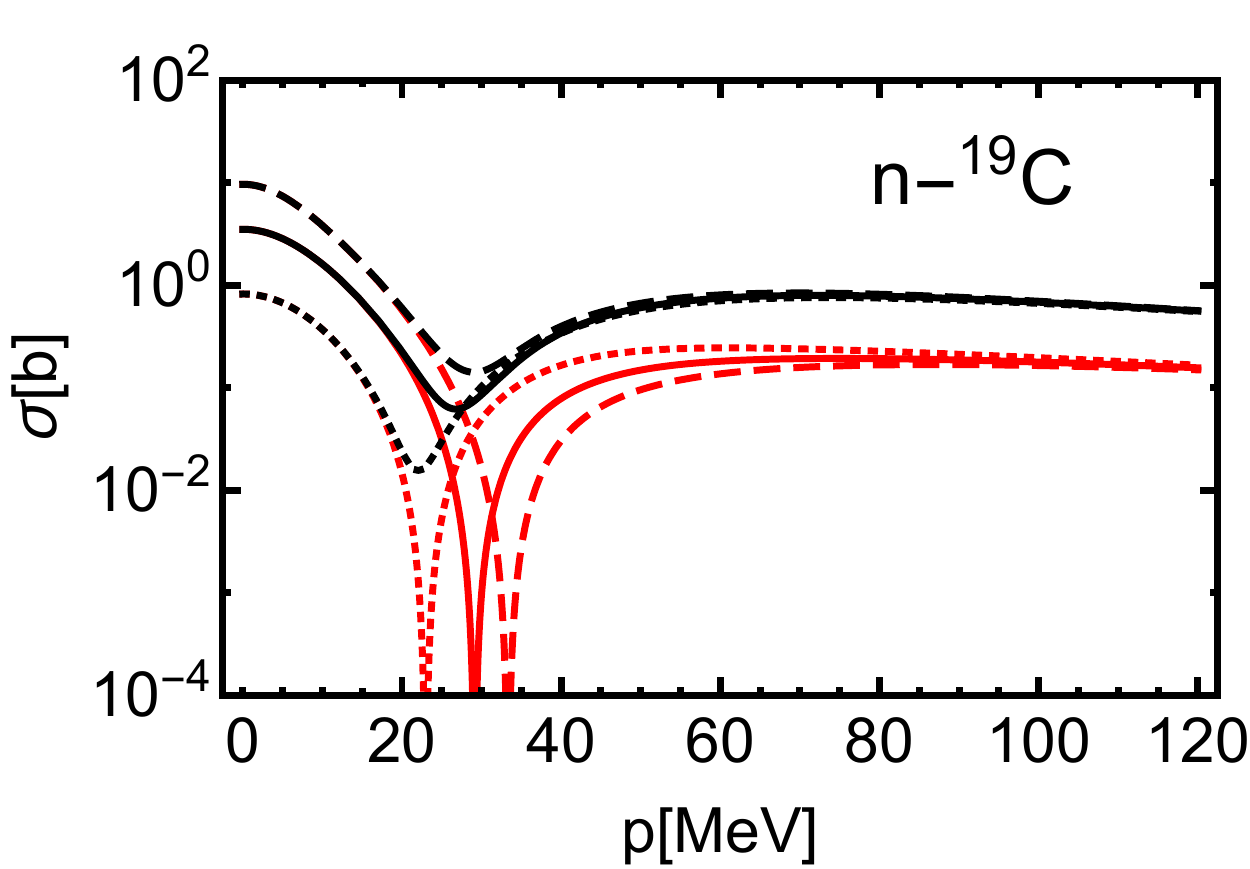}
\caption{Total cross sections considering only $s$-wave (red curves) and including higher partial waves with $l<5$ (black curves) of the neutron and one-neutron halo nuclei in the $J=0$ channel: $n\text{-}^{11}\rm Be$ (left), $n\text{-}^{15}\rm C$ (middle), and $n\text{-}^{19}\rm C$ (right). 
The notations of the solid, dashed and dotted curves are the same as those in Fig.~\ref{fig:kCotDeltaOne}. 
}
\label{fig:CrossSecTot}
\end{center}
\end{figure*}

Because of the existence of near-threshold bound states
in the $J=0$ channel for the $n$-$^{11}$Be, $n$-$^{15}$C and $n$-$^{19}$C systems in question, namely $^{12}$Be, $^{16}$C and $^{20}$C, respectively, and virtual excited states, the total cross sections shown as red curves in Fig.~\ref{fig:CrossSecTot} have much more prominent threshold enhancement in comparison with those in the $J=1$ channel shown in Fig.~\ref{fig:CrossSecOne}.
The obtained total cross sections in the near-threshold region are of the order of a few barns for the $n$-$^{11}\rm Be$ and $n$-$^{19}\rm C$ scattering and less than 1 barn for the $n$-$^{15}\rm C$ scattering. 
The threshold enhancement for the $n$-$^{15}\rm C$ scattering is milder than the other two cases because the $^{16}$C is more deeply bound than the $^{12}$Be and $^{20}$C (see the two-neutron separation energies in Table~\ref{Tab:InputB}).
One also sees that the pole of $p\cot\delta_0^R(p)$ shows up as a zero in the $s$-wave cross section for each of the considered processes; see the red curves in Fig.~\ref{fig:CrossSecTot}. We note that this zero in the cross section is an artefact of including only the contribution of $s$-wave scattering. If the contributions from higher partial waves are included these zeros will be filled up. However, as can be seen from the black curves in Fig.~\ref{fig:CrossSecTot}, which include contributions with $l < 5$, they remain visible as pronounced minima in the total cross section.
The observation of such minima at the predicted positions would provide clear evidence of the Efimov effect in halo nuclei.

Analog phenomena exist in other systems.
In a recent study of the double-charm tetraquark $T_{cc}$ in the $DD^*$ scattering with the $DD\pi$ three-body dynamics, a similar pole of 
$p\cot\delta_0^R(p)$ was also found~\cite{Du:2023hlu}.
In fact, the existence of a similar pole in $p\cot\delta_0^R(p)$ for the neutron-deuteron scattering in the triton channel was already observed more than half a century ago~\cite{Phillips:1969hm,Reiner:1969mmv}.
An effective field theory treatment of the excited Efimov state in the triton was presented in Ref.~\cite{Rupak:2018gnc}.

\section{Summary}\label{sec:Sum}

In recent years, various halo nuclei  with a tightly bound core surrounded by weakly bound valence nucleon(s) have been found. Exploring
the structure and reactions of halo nuclei will help us understand fundamental aspects of nuclear forces and nuclei at the edge of stability.
 
 In the present work, the $s$-wave interactions of neutron and spin-parity $J^P=\frac{1}{2}^+$ one-neutron halo nuclei 
$^{11}\rm Be$, $^{15}\rm C$ and $^{19}\rm C$ are studied using Halo EFT at LO. In the total spin $J=1$ channel, 
the only input to the Faddeev equation is the one-neutron separation energy for each one-neutron halo nucleus.
The total cross sections at threshold are of the order of a few barns for all the considered $n$-$^{11}\rm Be$, $n$-$^{15}\rm C$ and $n$-$^{19}\rm C$ scattering processes. 
In the total spin $J=0$ channel, the amplitudes of the $n$-$^{11}\rm Be$, $n$-$^{15}\rm C$ and $n$-$^{19}\rm C$ scatterings do not have a unique solution as the cutoff $\Lambda\to \infty$. Following Ref.~\cite{Bedaque:1998kg} a three-body counterterm $D_0$ is introduced to absorb the cutoff dependence and to achieve a unique solution of the Faddeev equation. The $D_0$ values are tuned to reproduce the $^{12}\rm Be$,  $^{16}\rm C$ and $^{20}\rm C$ ground state two-neutron separation energies. The numerical results show that the total $s$-wave cross sections at threshold are of the order of a few to ten barns for the $n$-$^{11}\rm Be$ and $n$-$^{19}\rm C$ scattering, and is much smaller ($\approx60$~mb) for the $n$-$^{15}\rm C$ scattering. This is because is the latter case, the $^{16}$C is more deeply bound than the other two cases and thus its enhancement effect at threshold is smaller. 

We also find that for the neutron-core scattering length in a certain range, the $s$-wave neutron-halo-nucleus scattering amplitude has a zero, corresponding to a pole of $p\cot\delta_0^R$ near threshold on the real positive $p$ axis. 
The location of the $T$-matrix zero depends on the neutron-core scattering length. 
This zero is manifestation of the presence of an excited virtual Efimov state close to the scattering threshold. Our results for $n$-$^{19}\rm C$ scattering are in qualitative agreement with the work of Refs.~\cite{Yamashita:2008sg} in the renormalized zero-range model. The qualitative features are unchanged by 
finite-range effects~\cite{Shalchi:2016psb,Deltuva:2017zvk}. In contrast to 
Refs.~\cite{Mazumdar:2006tn,Mazumdar:2011zz}, we find no evidence for a scattering resonance. 

Up to higher order corrections, the position of the pole of $p\cot\delta_0^R$ is fully determined by the separation energies of the corresponding one- and two-neutron halo nuclei and the neutron-neutron scattering length.
If it can be observed, {\it e.g.} through a
minimum in the scattering cross section,
it could serve as an experimental confirmation of Efimov physics in halo nuclei, even if bound excited states are absent. Thus it provides an alternative to previous proposals to observe the Efimov effect in halo nuclei~\cite{Macchiavelli:2015xml} and the standard approach
to observe excited states that satisfy the universal scaling relations~\cite{Federov:1994cf,Amorim:1997mq,Canham:2008jd,Hammer:2017tjm,Bishop:2020sqi}. While there is evidence that finite-range effects do not remove the minima~\cite{Shalchi:2016psb,Deltuva:2017zvk}, the calculation of higher-order corrections to their position would be valuable. Moreover, it would be interesting to elucidate the effects of virtual Efimov states in two-neutron halo nuclei in the scattering of the corresponding one-neutron halo nuclei off deuteron targets, which is directly accessible in experiment.

\acknowledgments 
We are grateful to Han-Tao Jing and Shan-Gui Zhou for helpful discussions.
This work is supported in part by the National Natural Science Foundation of China (NSFC)  under Grants No. 12247139, No. 12125507, No. 11835015, and No. 12047503,
by the Chinese Academy of Sciences (CAS) under Grant No.~YSBR-101 and No. XDB34030000; by the NSFC and the Deutsche Forschungsgemeinschaft (DFG, German Research Foundation) through the funds provided to the Sino-German Collaborative Research Center “Symmetries and the Emergence of Structure in QCD” (NSFC Grant No. 12070131001, DFG Project-ID 196253076 -  TRR110),
by the DFG under Project-ID 279384907 - SFB 1245, and by the German
Federal Ministry of Education and Research (BMBF) (Grant No. 05P21RDFNB).

\appendix

\section{Legendre polynomials of the second kind with complex argument}
\label{sec:legendre}
The Legendre polynomials of the second kind with complex argument are 
\begin{align}
&Q_0(z)=\frac{1}{2}\ln\frac{z+1}{z-1}\,, \\
&Q_1(z)=\frac{z}{2}\ln\frac{z+1}{z-1}-1\,,\\
&Q_2(z)=\frac{1}{4}(3z^2-1)\ln\frac{z+1}{z-1}-\frac{3}{2}z\,,\\
&Q_3(z)=\frac{1}{4}(5z^3-3z)\ln\frac{z+1}{z-1}-\frac{5}{2}z^2+\frac{2}{3}\,,\\
&Q_4(z)=\frac{1}{16}(35z^4-30z^2+3)\ln\frac{z+1}{z-1}-\frac{35}{8}z^3+\frac{55}{24}z\,.
\end{align}

\section{Numerical solution method}
\label{sec:numerical}

The Faddeev equation can be solved using the matrix inversion method~\cite{Haftel:1970zz}. 
When the total energy $E$ is above the two-body $n\sigma$ threshold, the $\sigma$ dimer can be on-shell. And the logarithmic singularities in the $s$-wave projection core-exchange and neutron-exchange potentials in Eqs.~\eqref{Eq:CEP} and \eqref{Eq:NEP} will also be in the integration region when
$E$ is above the three-body $nnc$ threshold. Both of these will cause numerical instabilities as the integration in Eqs.~\eqref{Eq:FaddSpinOne} and \eqref{Eq:FaddSpinZero} is performed along the positive real momentum axis. 
To overcome that problem, we use the contour deformation method~\cite{HetheringtonSchick,Aaron:1966zz,Cahill:1971ddy,BookThree}. The basic idea is that the integration contour in Eqs.~\eqref{Eq:FaddSpinOne} and \eqref{Eq:FaddSpinZero} can be distorted from its original position along the positive real axis into
the complex momentum plane without changing the results. One should note that the angle of the deformed contour should be large enough in order to avoid the singularities on the real axis but small enough in order not to cross the singularities in the
core-exchange and neutron-exchange potentials. Using the Cauchy's theorem, the relation between the solution of the Faddeev equation for real momentum and complex momentum can then be obtained.

To be concrete, we take the single channel case with $J=1$ as an example, and the extension to the coupled-channel case is straightforward.
First, the Faddeev equation is analytically continued into the complex momentum plane using Cauchy's theorem,
 \begin{widetext}
\begin{equation}
\label{Eq:FaddComplex}
iT_{\sigma\sigma}(k',p,E)=iV_{\sigma\sigma}(k',p,E)-\int_C\frac{q'^2dq'}{2\pi^2}V_{\sigma\sigma}(k',q',E)Z_{\sigma}^{-1}D_{\sigma}\!\left(E-\frac{q'^2}{2m_n},q'\right)iT_{\sigma\sigma}(q',p,E)\,,
\end{equation}
where $k'$ and $q'$ are complex momenta. And now the integration is along the deformed contour $C$, which will be discussed later. 
We can see that the Faddeev equation in Eq.~\eqref{Eq:FaddComplex} is free of singularity for $E$ on the real axis, and this equation can be solved 
using the matrix inversion method without any difficulty. The $T$-matrix for real momentum $k$ can then be obtained. 
\end{widetext}

Since we are interested in the on-shell $T$-matrix, we put the outgoing momentum $k$ on the energy shell, which leads to
\begin{equation}
k=\left[2\mu_{n\sigma}\left(E+\frac{1}{2\mu_{nc}a_{\sigma}^2}\right)\right]^{1/2}\,,
\end{equation}
where $\mu_{n\sigma}$ is defined in Eq.~(\ref{Eq:nsigmared}).
Inserting this into Eqs.~\eqref{Eq:CEP} gives the logarithmic branch points of $q$ satisfying
\begin{equation}
E-m_\sigma/(2m_nm_c)(q^2+k^2) \pm qk/m_c+i\epsilon =0,
\end{equation}
and the solutions of this equation give the locations of four branch points of $q$ in the complex momentum plane,
\begin{equation}
q=\frac{\pm m_n k \pm im_{\sigma}/a_{\sigma}}{m_{\sigma}}\,.
\end{equation}
In addition to these branch points, the dimer propagator has a pole at 
\begin{equation}
q=\left[2\mu_{n\sigma}\left(E+\frac{1}{2\mu_{nc}a_{\sigma}^2}+i\epsilon\right)\right]^{1/2} .
\end{equation}

\begin{figure}[t]
\begin{center}
\includegraphics [scale=0.43] {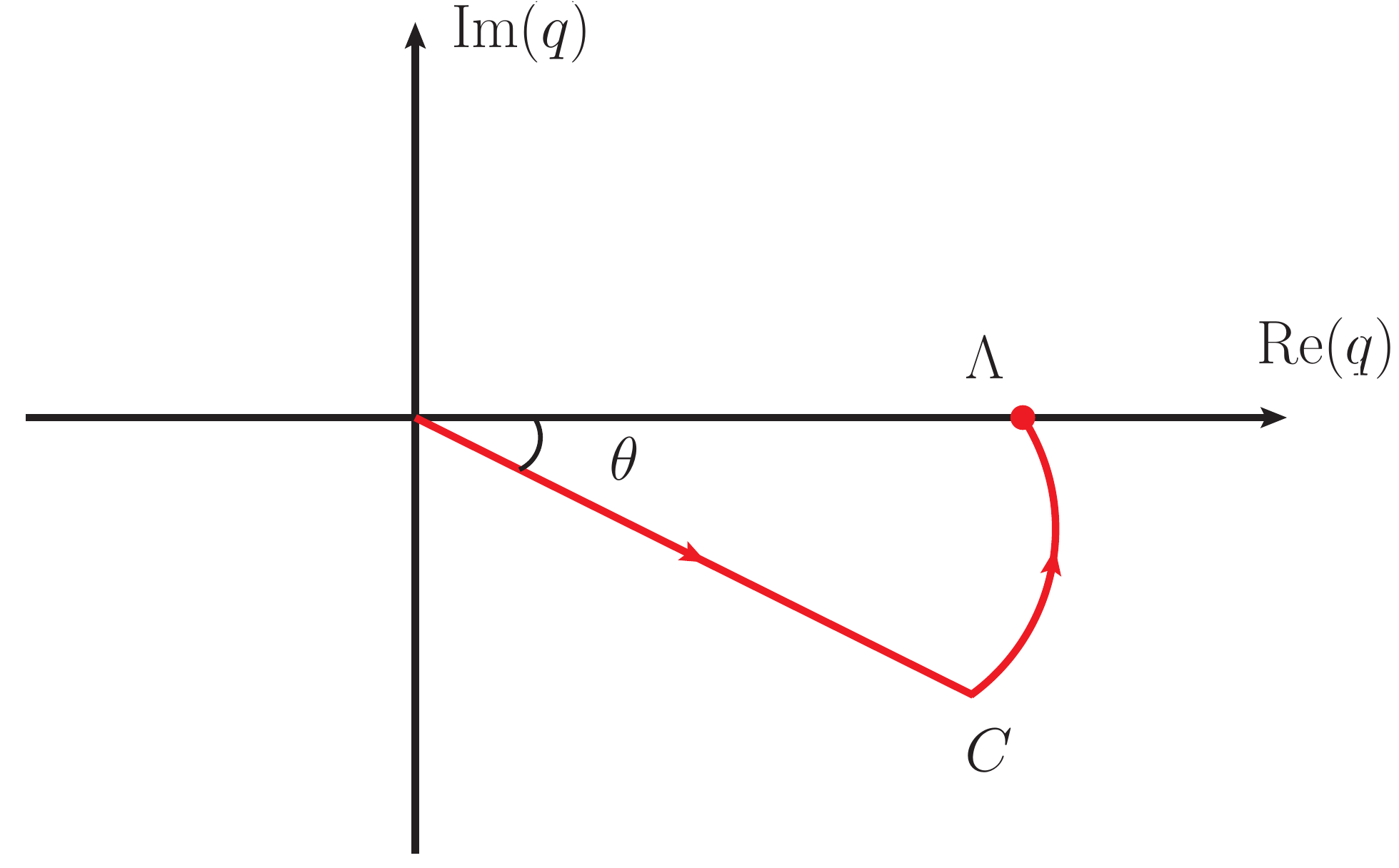}
\caption{The integration contour $C$ for the evaluation of Eq.~\eqref{Eq:FaddComplex}.}
\label{fig:Conto}
\end{center}
\end{figure}

In our calculations, we rotate the integration contour into the lower half $q$-plane, and $q'$ is along the contour $C$ as shown in Fig.~\ref{fig:Conto} 
to avoid singularities in the dimer propagator and the core-exchange potential in Eq.~\eqref{Eq:FaddComplex}. 
The locations of the logarithmic branch points place an upper limit on $\theta$, which is the angle between the clockwise-rotated integration path and the real $q$ axis,
\begin{equation}
0<\theta < {\rm {arctan}}\left( \frac{m_{\sigma}/a_{\sigma}}{m_n k}\right) .
\end{equation}
We have checked that the numerical results are independent of the choice for $\theta$ as long as its value satisfies the above constraint.

\bibliography{ref.bib}
 
\end{document}